\documentclass[11pt]{article}   	
\usepackage{geometry}                		
\usepackage{graphicx}				
\usepackage{amsfonts}
\usepackage{amssymb}
\usepackage{tikz-cd}
\usepackage{nicefrac}
\usepackage{etoolbox}
\usepackage{pdflscape}
\usepackage{rotating}
\usepackage{bm}
\usepackage{diagbox}
\usepackage{algpseudocode}
\usepackage{nicematrix}
\ExplSyntaxOn
\makeatletter
\tl_set:Nn \l_tmpa_tl {pgfsys-dvips.def} \tl_if_eq:NNT \l_tmpa_tl \pgfsysdriver
{\cs_set:Npn\pgfsys@blend@mode#1{\special{ps:~/\tl_upper_case:n #1~.setblendmode}}} \makeatother
\ExplSyntaxOff
\usepackage{cleveref}
\usepackage{dcolumn}
\newcolumntype{.}{D{.}{.}{3}}

\algnewcommand{\IIf}[1]{\State\algorithmicif\ #1\ \algorithmicthen}
\algnewcommand{\EndIIf}{\unskip\ \algorithmicend\ \algorithmicif}
\algnewcommand{\FFor}[1]{\State\algorithmicfor\ #1\ \algorithmicdo}
\algnewcommand{\EndFFor}{\unskip\ \algorithmicend\ \algorithmicfor}

\renewcommand{\vec}[1]{\boldsymbol{#1}} 

\newcommand{\loc}{\ensuremath{\lambda}}
\newcommand{\wid}{\ensuremath{\varepsilon}}
\newcommand{\weight}{\ensuremath{w}}

\newcommand{\param}{\ensuremath{t}}
\newcommand{\eparam}{\ensuremath{\tau}}
\newcommand{\rate}{\ensuremath{\rho}}

\newcommand{\ratediff}{\ensuremath{\Delta}}

\newcommand{\wtx}{\ensuremath{\widetilde{x}}}
\newcommand{\wtwid}{\ensuremath{\widetilde{\wid}}}
\newcommand{\wtloc}{\ensuremath{\widetilde{\loc}}}

\newcommand{\numer}{\ensuremath{m}}
\newcommand{\denom}{\ensuremath{n}}

\newcommand{\clause}{L}
\newcommand{\depth}{\ensuremath{D}}

\newcommand{\Poly}{\ensuremath{{\cal P}}}
\newcommand{\Sat}{\ensuremath{{\Xi}}}
\newcommand{\Tay}{\ensuremath{{Y}}}
\newcommand{\Bez}{\ensuremath{{Z}}}

\newcommand{\matsep}{4pt}

\newcolumntype{d}{D{.}{.}{1.3}}

\title{Perturbative methods for mostly monotonic\\ probabilistic satisfiability problems}
\date{}							
\author{Stephen Eubank\thanks{Biocomplexity Institute, University of Virginia 
  ({eubank@virginia.edu}, {abhijin@virginia.edu}).}
\and Madhurima Nath\thanks{Slalom,  
  ({mnath@vt.edu}).}
\and Yihui Ren\thanks{Brookhaven National Laboratory,  
  ({yren@bnl.gov}).}
\and Abhijin Adiga\footnotemark[1]}

\begin{document}

\maketitle

\begin{abstract}
The probabilistic satisfiability of a logical expression is a fundamental concept known as the partition function in statistical physics and field theory, an evaluation of a related graph's Tutte polynomial in mathematics, and the Moore-Shannon network reliability of that graph in engineering. 
It is the crucial element for decision-making under uncertainty.
Not surprisingly, it is provably hard to compute exactly or even to approximate.
Many of these applications are concerned only with a subset of problems for which the solutions are monotonic functions.
Here we extend the weak- and strong-coupling methods of statistical physics to heterogeneous satisfiability problems and introduce a novel approach to constructing lower and upper bounds on the approximation error for monotonic problems.
These bounds combine information from both perturbative analyses to produce bounds that are tight in the sense that they are saturated by some problem instance that is compatible with all the information contained in either approximation.
\end{abstract}

satisfiability, weak-coupling, strong-coupling, network reliability, complex networks

\section{Introduction}
At the heart of some of the thorniest problems in physics, computer science, engineering, and combinatorics lies a question that can be stated simply:
given the probabilities that any of a large set of events occur, what is the overall probability that a logical statement about combinations of those events is true?
We can think of the statement, ${\cal E}(c)$, as an assertion that a dynamical system in configuration $c$ has a certain property.
For example, given a lattice of magnets free to flip their orientations, what is the probability that two magnets separated by $k$ lattice sites point in the same direction? 
Or given the presence of an electron at position $x_0$ at time $t_0$, what is the probability of observing an electron at position $x_1$ at time $t_1$?
Or given the probability of failure for  ``crummy''  \cite{Moore:56} relays in an electrical circuit, what is the probability of establishing a current from one terminal to another?
This question is known as the probabilistic satisfiability\cite{finger2011probabilistic} (PSAT) problem.
Hard enough in the homogeneous case, when the probabilities of every event are the same, it becomes nightmarish in the more important heterogenous case, when the probability of each event can be different.
Heterogeneous systems arise if, for example, the lattice of magnets is stretched anisotropically, or each relay in the circuit is different.
A general method for answering it would find immediate technological applications in designing networks to avoid cascading failure or optimizing vaccination strategies, among many others.
Here we synthesize the explicit symmetries of Max Flow / Min Cut properties with the mathematical framework of network reliability and the perturbative analyses of statistical physics and field theory to produce a hierarchy of approximate solutions with controllable, bounded error.

A system's {\em configuration} is an assignment of states to each of its elements.
The system is defined by a probability distribution, $p$, over configurations.
In a system with a finite number of configurations, the probability that ${\cal E}$ is true is simply the sum of the probabilities of all configurations for which ${\cal E}$ is true, or, in terms of the indicator function $\delta$ which is 1 if its argument is true and 0 else, $p({\cal E}) = \sum_{c\in{\cal C}} \delta({\cal E} (c))
 p(c)$.
Partitioning the configurations into equiprobable equivalence classes, ${\cal K}$, this can be written as $p({\cal E}) = \sum_{k\in{\cal K}}  n(k) p_k $, where $n(k)$ is the fraction of configurations that have probability $p_k$ for which ${\cal E}$ is true.
The function $n(k)$ is known as the {\em density of states} function.

When only the relative probability $\widehat{p}$ of different configurations is known, properly normalized probabilities can be obtained as $p = \widehat{p}/Z$, where the normalization constant $Z =    \sum_{k\in{\cal K}} n(k) \widehat{p}_k $ is known as the {\em partition function}.
For a large class of physical systems, $p$ is the Boltzmann distribution, an exponential distribution that depends on the energy $E(c)$ of a configuration and the inverse of the system's temperature, $\beta$, i.e., $p(c) \propto e^{-\beta E(c)}$.
In this case, the normalization ``constant'' is a function of temperature: 
\begin{align}
Z(\beta) =  \sum_{k=1}^K n(k) e^{-\beta E_k},
\end{align}
assuming every configuration has one of $K$ distinct energies, $E_k$.
The partition function is key to understanding statistical systems because its logarithm is the moment generating function.
For example:
\begin{align}
-\frac{\partial}{\partial \beta} \ln Z &= \frac{ \sum_{k=1}^K E_k n(k) e^{-\beta E_k}}{Z} = \langle E \rangle.
\end{align}

Unfortunately, since evaluating the partition function is equivalent to solving PSAT or solving a network reliability problem, it is in the complexity class \#P \cite{valiant1979complexity}.
In practice, the solution is computationally infeasible for the foreseeable future even for fairly small sets of events.
Exact solution requires summing over each of $2^N$ possible configurations.
Fortunately, when ${\cal E}$ is {\em monotonic} -- i.e., the sense of each event can be chosen so that ${\cal E}$ 
contains no negations -- the oracle does not need to be called for every configuration. 
Typical expressions in physical systems exhibit regularities that allow some shortcuts, such as explicit forms for the  density of states, but this is not generally the case in other domains.

There are several approaches to approximating $Z$ for a monotonic ${\cal E}$ when $\widehat{p}$ is known:
\begin{enumerate}
\item {\em Probabilistic methods} and {\em simulations} approximate $n(k)$. They rely on algorithms that count approximately how many configurations are in each equivalence class in ${\cal K}$ and how many of those satisfy ${\cal E}$.
\item {\em Renormalization} produces an {\em effective} theory with fewer degrees of freedom and thus fewer configurations in the sum. 
Renormalization recursively ``integrates out'' some degrees of freedom -- e.g., alternate sites in a lattice -- to define a new problem with the same solution.
Indeed, PSAT is equivalent to an evaluation of a Tutte polynomial, which can be defined by recursive renormalization \cite{beaudin2010little}.
\item {\em Perturbative methods} compute leading terms in a Taylor series expansion of the solution for a system whose parameters have been perturbed slightly away from a system with a known solution.
\end{enumerate}
Roth has shown that approximately counting solutions is also hard  \cite{ROTH1996273}, but his arguments apply to {\em relative}, not absolute, error:
\begin{quotation}
The notion of approximation we [Roth] use is that of relative approximation. 
\ldots
For example, there exists a polynomial time randomized algorithm that approximates the number of satisfying assignments of a DNF formula within any constant ratio. 
It is possible, though, for a \#P-complete problem, even if its underlying decision problem is easy, to resist even an efficient approximate solution. \ldots We prove, for various propositional languages for which solving satisfiability is easy, that it is NP-hard to approximate the number of satisfying assignments even in a very weak sense.
\end{quotation}
Probabilistic satisfiability places the problem in a continuous rather than discrete context, where notions of smoothness make sense.
Perturbative approaches, called weak- and strong-coupling expansions in statistical physics, take advantage of smoothness to provide excellent approximations for many problems \cite{domb1949order}.
Indeed, perturbative approximation of the ``propagator'' in quantum electrodynamics (which is an infinite PSAT problem) resulted in what is generally acknowledged to be the most careful comparison between experiment and theoretical implications of a physical theory ever made  \cite{feynman2014qed}.

Here we develop practically useful approximations for finite, monotonic PSAT using a novel combination of perturbative and probabilistic methods with some elements of renormalization and simulation.
Furthermore, we show how to interpolate between two perturbative approximations to enforce constraints on unitarity and monotonicity that the truncated Taylor series do not obey.
Finally, we use the interpolated perturbation series to construct bounds on the approximation error, which improve on bounds constructed using only one or the other \cite{krivoulets2001theory}.
The interpolation relies on three principles -- positivity, unitarity, and duality -- that emphasize different aspects of the simple observation that all probabilities lie in the interval $[0,1]$.
The approach provides a nested hierarchy of perturbative approximations with bounded error giving a controllable trade-off between computational complexity and approximation error.
The approximations are controlled by two parameters: $S$, the number of samples generated in the probabilistic part, and $\depth$, the depth of expansion in the perturbative part. 
For large enough $S$ and $\depth$, the approximation becomes exact.
When $\depth$ is small, the quality of approximation depends on the problem instance.
When $S$ is sufficiently large, the bounds on approximation error are tight in the sense that there are expressions ${\cal E}'$ that are consistent with the perturbative approximations and saturate the bounds.
When $S$ is small, the bounds are not strict for ${\cal E}$, but they are tight bounds on a simpler problem ${\cal E}'$ that contains all the information about ${\cal E}$ that is available in the sample.

\Cref{sec:statement} gives a formal statement of the problem.
In particular, \cref{sec:approximation} reconciles the claims for our approximations with the fact that even approximating the satisfiability is NP-hard.
\Cref{sec:graph} maps the problem to a graphical setting analogous to Feynman diagrams, where the satisfiability becomes the propagator, the 2-terminal Moore-Shannon network reliability \cite{Moore:56} or, equivalently, an evaluation of the Tutte polynomial.
The approximation methods are described in \cref{sec:methods} for the case of identically distributed variables.
A fully worked, nontrivial example is explained in \cref{sec:example}.
The extension to non-identically distributed variables is developed in \cref{sec:heterogeneous} and applied to several variants of the example in \cref{sec:example2}.

\section{Problem statement}
\label{sec:statement}

\subsection{Events}
\label{sec:events}
Consider a set of $N$ events $E_i$ which might or might not occur in any instance of a random process.
We assume the events occur independently of each other.
Sometimes the events must be defined carefully to ensure they are truly independent.
For example, in simple models of disease transmission, whether one person is infected depends on who else is infected, whereas whether one infected person transmits to a susceptible person is independent of other transmission events.
The former is the kind of compound event that is captured by the expression ${\cal E}$, but it is not an atomic event $E$.
With each event $E_i$ for $i \in \{1,\ldots,N\}$, we associate a Bernoulli random variable $e_i \in \{0,1\}$ and a probability $\wtx_i\in[0,1]$ that $e_i = 1$, meaning that $E_i$ occurred.
Until \cref{sec:heterogeneous}, we assume a homogeneous system with $\wtx_i = x$ for all $i$. 

\subsection{Satisfiability Expression}
\label{sec:expression}

Suppose there is a monotonic Boolean expression ${\cal E}$ in disjunctive normal form, i.e.,
\begin{equation}
{\cal E} = \bigvee_{i=1}^m \left(\bigwedge_{j\in \clause_i} e_j\right),
\label{eq:dnf}
\end{equation}
where $\clause_i \subseteq \{1,\ldots,N\}$.
To avoid trivial cases we assume that $|\clause_i| \ge 1$ and $m \ge 1$.
Given an assignment of values to every $e_i$ -- i.e., a {\em system configuration} -- the expression ${\cal E}$ evaluates deterministically to either true or false.
We say the expression is {\em satisfied} if it evaluates to true, and the configuration is a {\em solution}.
We refer to each of the $m$ conjunctions in Equation~\cref{eq:dnf} as a {\em clause}.
Each clause is defined by a set $\clause_i$ of events that must all occur in order for the clause to be satisfied.
We also refer to the set of {\em integers} $\clause_i$ as {\em clause} $i$ when it causes no ambiguity.

\subsection{Satisfiability Problems}
The deterministic satisfiability literature focuses on counting solutions.
{\em Probabilistic} satisfiability 
asks instead:
What is $\Sat({\cal E}, \vec{\wtx})$, the probability that the expression ${\cal E}$ is satisfied given the probability of individual events $\wtx_i$? 
 $\Sat({\cal E}, \vec{\wtx})$ is a sum of probabilities over all solutions;
evaluating it is thus at least as hard as counting the solutions. 
Indeed, when every event occurs with probability $\nicefrac{1}{2}$, the satisfiability reduces to the ratio of the number of solutions to the number of configurations, $2^N$.

\subsection{On the nature of approximation}
\label{sec:approximation}

\begin{figure}[htbp]
\begin{center}
\includegraphics[width=0.32\textwidth]{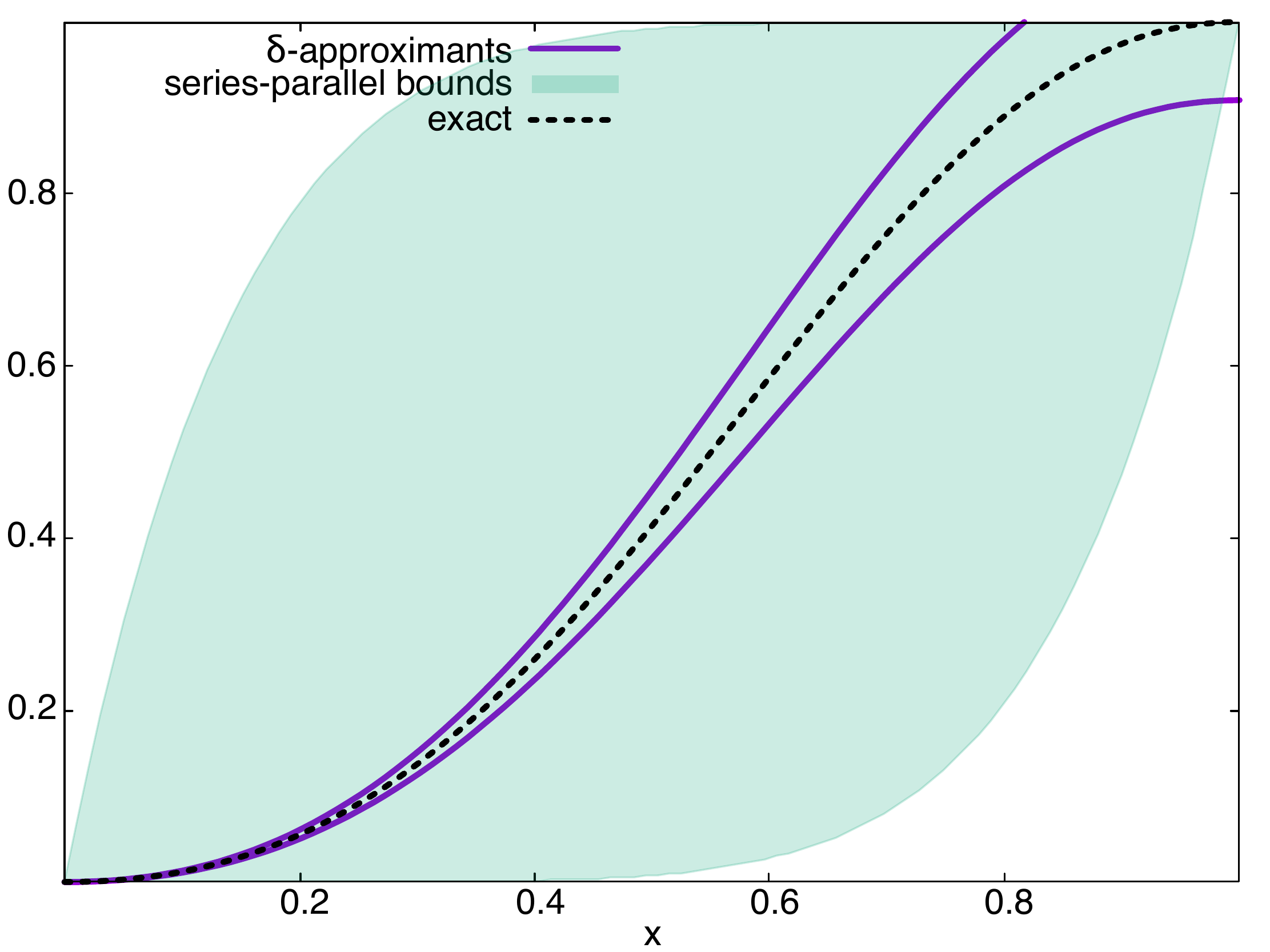}\hfil
\includegraphics[width=0.32\textwidth]{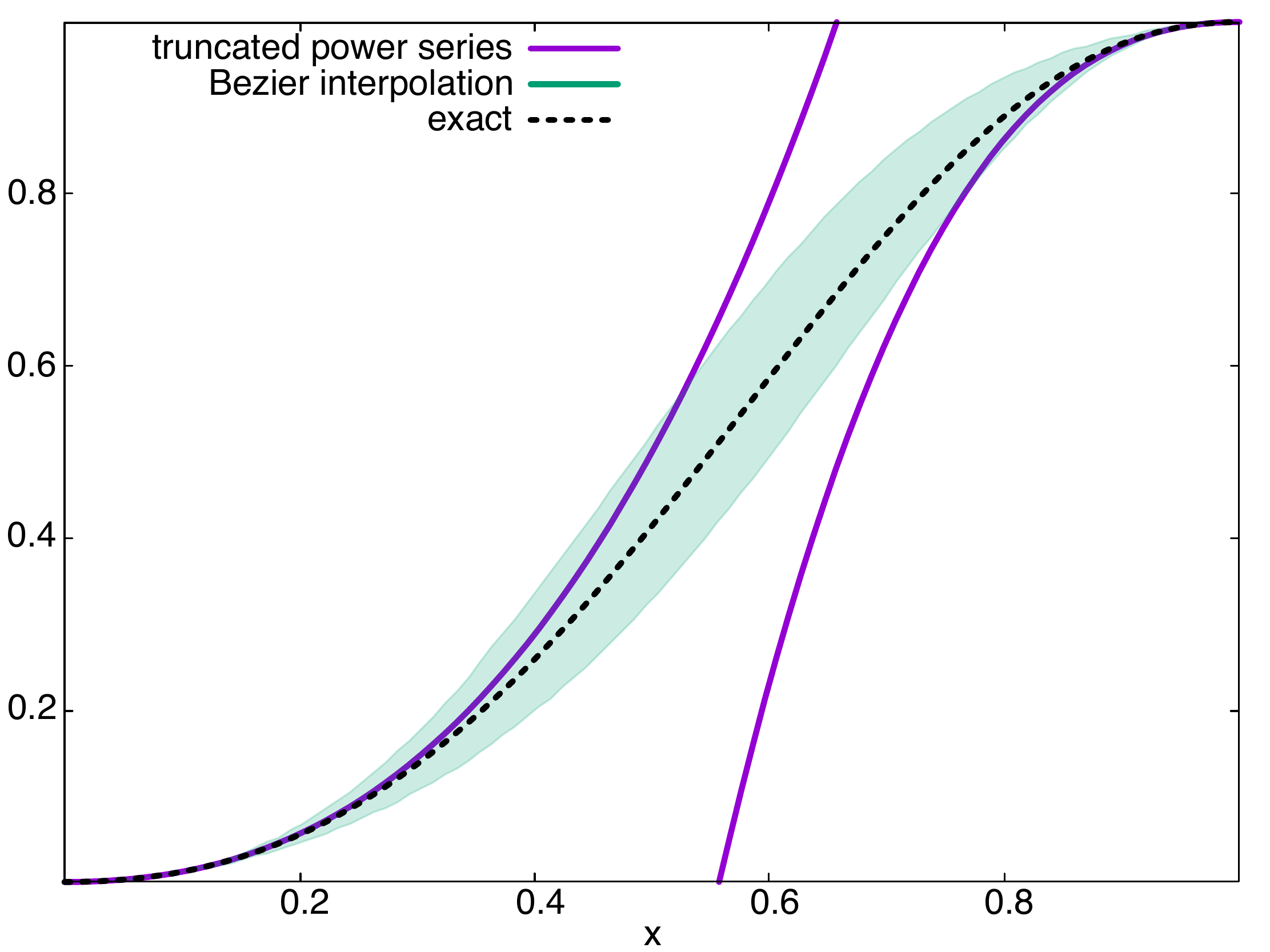}\hfil
\includegraphics[width=0.32\textwidth]{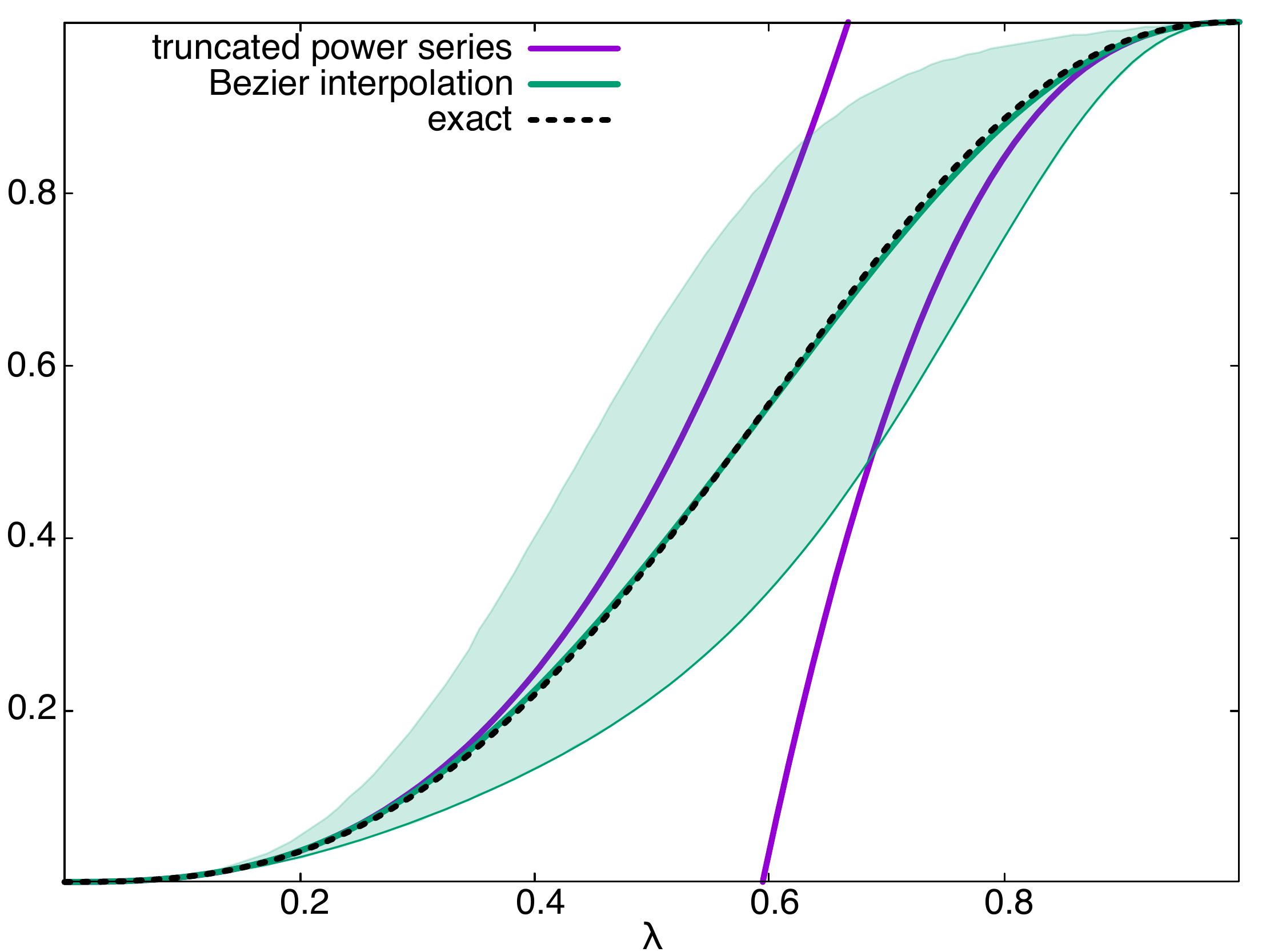}
\caption{Satisfiability for the example expression in \cref{eq:exampleE}, as a function of $x =p(e_2)$.
(Left) Upper and lower bounds on (hypothetical) $\delta$-approximants to the satisfiability, with $\delta = 0.1$, along with the strict, but not tight, bounds of \cref{eq:seriesParallel} based only on the number of events in the expression.
(Center) Truncated Taylor series for the homogeneous example to  $O(x^3)$ at $x=0$ and $O((1-x)^2)$ at $x=1$, along with the bounds and interpolation developed here. 
(Right) The same as the center panel, but for the heterogeneous problem described in \cref{sec:example2} with $(\wtx_1,\wtx_6)=(\nicefrac{1}{2},\nicefrac{1}{4})$.
}
\label{fig:example}
\end{center}
\end{figure}

It is important to distinguish between the notions of ``approximation'' used here and in algorithmic complexity.
Specifically, Roth considers both {\em relative approximation} -- the result $M'$ is a $\delta$-approximation to  $M$ if and only if $M'/(1+\delta) \le M \le M' (1+\delta)$ for $\delta \ge 0$ -- and the use of
{\em approximate theories} -- expressions ${\cal E}'$ and ${\cal E}''$  for which $p({\cal E}') \le p({\cal E})  \le p({\cal E}'') $.
Relative approximation is known to be NP-hard for PSAT, but approximate renormalization and perturbative methods produce approximate theories.
The left panel of \cref{fig:example} illustrates differences in the character of results for the example in \cref{sec:example,sec:example2}.
The simplest perturbative approximations, based only on the number of events in ${\cal E}$ as in \cref{eq:seriesParallel}, constrain the exact solution to lie in the green band in the figure;
a relative approximation to the exact answer (dashed curve) with $\delta=0.1$ would lie between the two solid curves.
Clearly, at this order, the perturbative approximation is almost useless, and Roth correctly judged the use of approximate theories to be ineffective for the decision problem he studied.
However, notice that relative approximation does not automatically incorporate constraints such as unitarity or duality, and that, of course, the absolute error can be large where the exact answer is large.
By including more information about the expression in a perturbative analysis, it is possible to ``squeeze'' the bounds dramatically, as illustrated in the center and right panels of the figure.
The resulting approximation respects unitarity and duality, and provides small absolute errors at both ends of the domain.
However, the bounds cannot be squeezed much in the middle, where, as we show below, there is little information, so the relative error cannot be reduced uniformly below an arbitrary threshold $\delta$.

\subsection{Monotonicity, minimality, and total probability}
\label{sec:configs}

\begin{figure}[htbp]
\begin{center}
\includegraphics[width=0.32\textwidth]{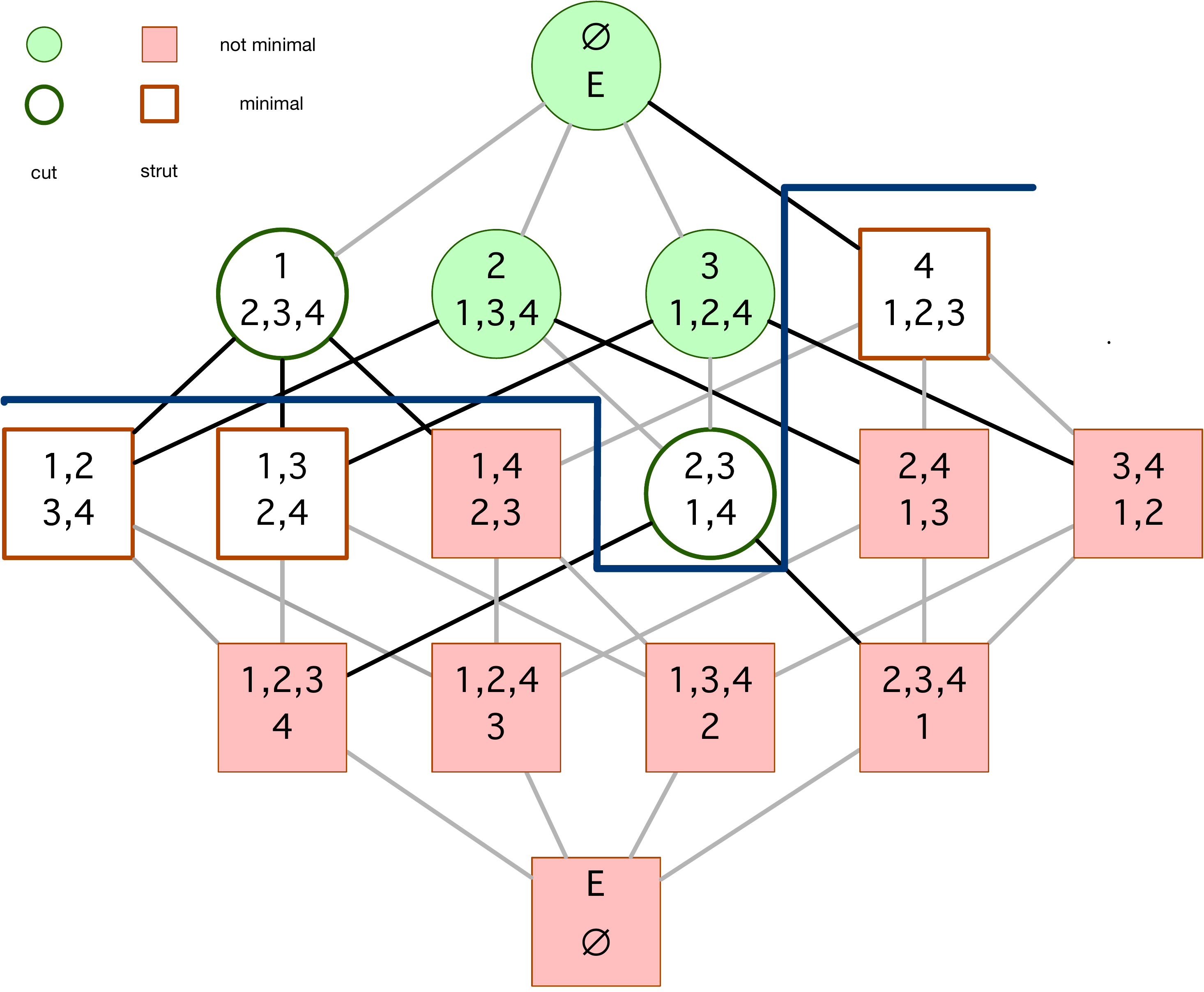}
\hfil
\includegraphics[width=0.32\textwidth]{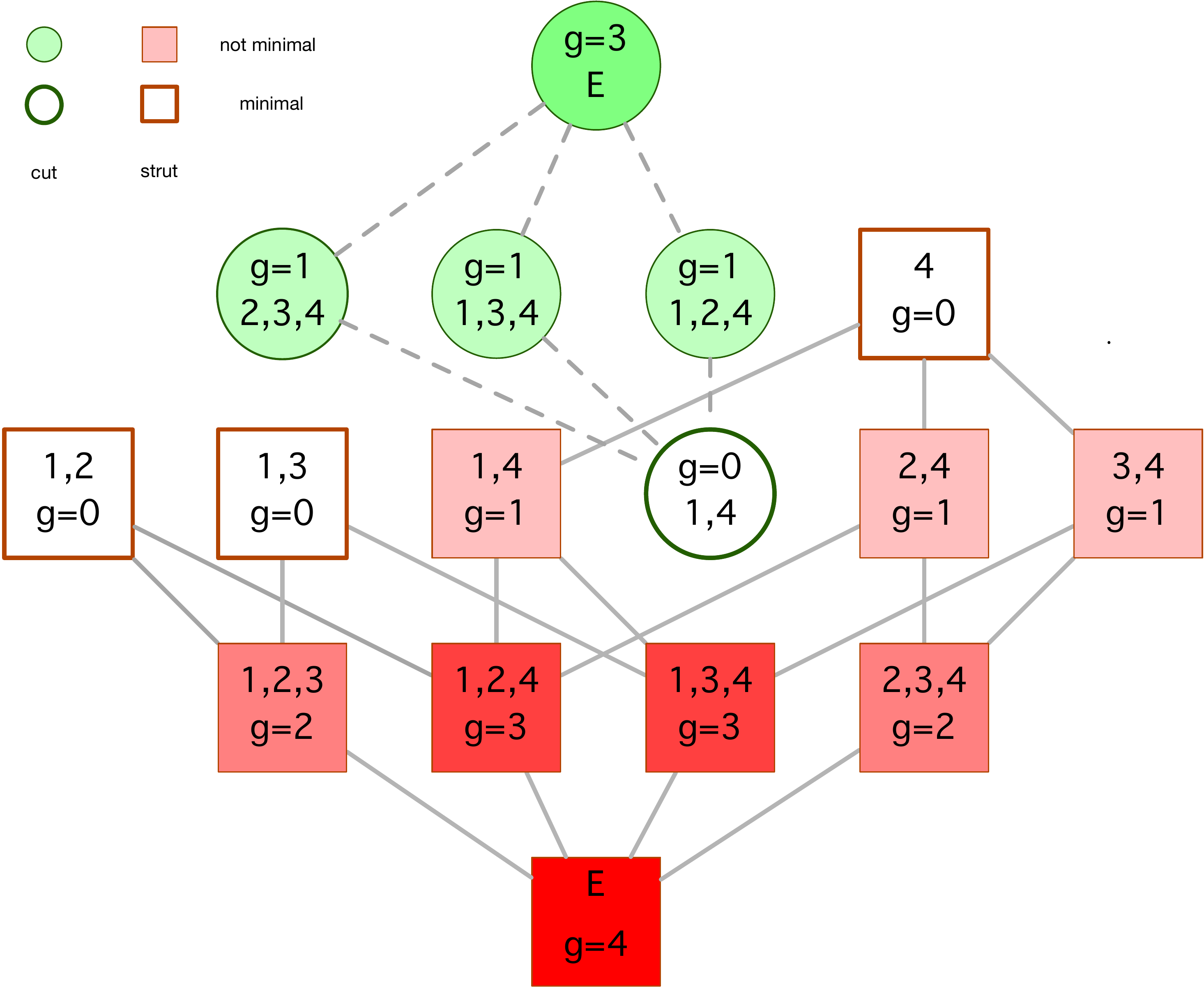}
\hfil
\includegraphics[width=0.25\textwidth]{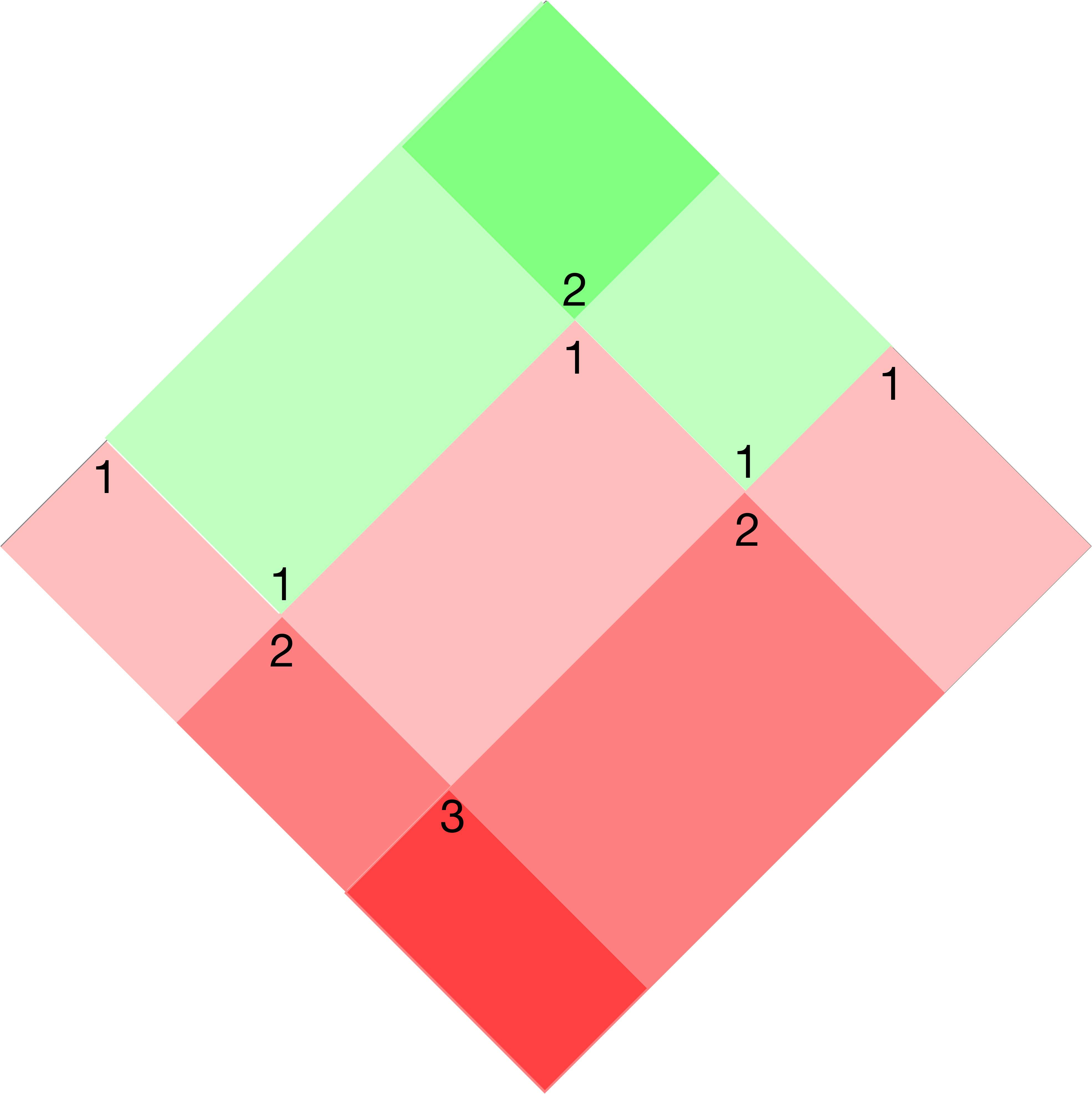}
\caption{Perspectives on the partially ordered sets corresponding to the expression ${\cal E} = (e_1 \land e_2) \lor (e_1 \land e_3) \lor e_4$. Each configuration is labelled (redundantly) with the indices of the events with $e_i=1$ on the top line and those with $e_i=0$ on the bottom. The $i$-th row, starting from $i=0$ at the top, contains all configurations with exactly $i$ true events. 
Cuts are indicated by circles; struts by squares. Filled circles or squares are not minimal. 
Left: A separatrix between cuts and struts in the lattice of all possible configurations for a monotonic expression over $N=4$ events.  Darker edges indicate those that connect a strut to a cut. The separatrix is the thick, orthogonal curve from the left to the right edge of the lattice. All minimal struts and cuts are adjacent to the separatrix, but the converse is not true.
Center: The descendants of minimal cuts and struts and the degeneracy of each configuration (the number of direct ancestors that are cuts or struts). Only the edges that do not cross the separatrix are shown, for clarity.
Right:  A notional depiction of the ``light cones'' of minimal struts (red) and cuts (green). The degeneracy of configurations within a cone is indicated by the color's shade and labeled at the minimum configuration of the region.
If there are more than two minimal struts, the partition into cones is not necessarily planar, as illustrated in the center panel.
Here, the minimal struts are either on the boundary or adjacent to cuts with degeneracy 2 and vice versa.
}
\label{fig:separatrix}
\end{center}
\end{figure}

The set of all configurations is the {\em power set} of the set of events.
Any configuration can be labeled by the set of events for which $e_i=1$.
It is convenient to impose the partial order of set inclusion on the set of all configurations and to represent the graded partial order, with the rank of configuration labeled $s$ given by $|s|$, as the lattice depicted in \cref{fig:separatrix}. 
We define $D(c)$ to be the set of all descendants of $c$ in the lattice.
That is, $D(c) \equiv \{c' \in {\cal C} | c' \supseteq c\}$.

The {\em total probability} $p_t$ of a configuration $c$ is the probability that the associated clause is true, i.e., the sum of probabilities of all its descendants that satisfy ${\cal E}$.
Clearly, $p_t(c) \ge p(c)$.
For a monotonic system, if $c$ satisfies ${\cal E}$, so do all its descendants.
If the events are independent of each other, the probability of a configuration factorizes into the product of probabilities of the individual events.
Thus, for a configuration $c$ that satisfies ${\cal E}$,
\begin{align}
p_t(c) &= p\left(\bigwedge_{i\in c} e_i\right) = \sum_{c'\in D(c)} p(c') \delta({\cal E} | c') \\
&\xrightarrow{monotonic} \sum_{c'\in D(c)} p(c')
\xrightarrow{independent} \prod_{i\in c} \wtx_i \xrightarrow{homogeneous} x^{|c|}.
\label[equation]{eq:probClosure}
\end{align}
Because the descendants of any two configurations $c_1$ and $c_2$ are not disjoint, the probability of a disjunction of clauses is sub-additive.
The Inclusion-Exclusion principle gives
\begin{equation}
p(\clause_i \lor \clause_j) = p_t(\clause_i) + p_t(\clause_j) - p_t(\clause_i \cup \clause_j)
\label{eq:IEstep}
\end{equation}

A finite, monotonic system admits  the notion of {\em minimal} solutions.
A configuration $c$ is a minimal solution if it is a solution, but no proper subset of $c$ is a solution.
See examples in \cref{sec:example}.
Without loss of generality, we will assume that each clause in ${\cal E}$ is a different minimal solution, i.e., for distinct $i$ and $j$, $\clause_i \nsubseteq \clause_j$.
Under this condition, 
\begin{equation}
\sum_{c \in {\cal L}} p(c) \le p({\cal E}) \le \sum_{c \in {\cal L}} p_t(c).
\label{eq:simpleBounds}
\end{equation}
These simplistic bounds could be tightened recursively, but it will be easier to use the duality introduced in \cref{sec:dual}.

The descendants of $\clause_i \cup \clause_j$ are all included in both the first and second terms of  \cref{eq:IEstep}; the third term serves to correct this overcounting.
In general, we define a configuration's {\em degeneracy} as the number of distinct minimal solutions it has among its ancestors, including itself.
The right panel of \cref{fig:separatrix} illustrates the degeneracy in a simple example.
Just as there is a set of minimal solutions in a monotonic problem, there are sets of minimal configurations with degeneracy $1 \le k \le L$, which we denote by ${\cal M}_k$.
(Notice that ${\cal M}_1 = {\cal L}$.)
One way to construct ${\cal M}_k$ is first to construct all possible unions of exactly $k$ elements of ${\cal M}_1$, then to remove those that are supersets of others.
Iterating the Inclusion-Exclusion principle in \cref{eq:IEstep}, we can reorganize the sum into a form that produces a power series: 
\begin{equation}
p({\cal E}) = \sum_{c\in{\cal C}} \delta({\cal E}|c) p(c) = \sum_{k=1}^{L} (-1)^{k+1}  \sum_{c\in{\cal M}_k} p_t(c) \xrightarrow[homogeneous]{independent}\sum_{k=1}^{L} (-1)^{k+1}  \sum_{c\in{\cal M}_k} x^{|c|}.
\end{equation}

\subsection{Duality, struts, and cuts}
\label{sec:dual}

The logical complement of the expression ${\cal E}$, denoted $\overline{{\cal E}}$, can be expressed in disjunctive  normal form using the complements of each random variable, $\overline{e}$:
\begin{equation}
\overline{{\cal E}} = \overline{\bigvee_{i=1}^m \left(\bigwedge_{j\in \clause_i} e_j\right)} 
=\bigwedge_{i=1}^m \left(\bigvee_{j\in \clause_i} \overline{e}_j\right)
 = \bigvee_{i=1}^{\overline{m}} \left(\bigwedge_{j\in \overline{\clause}_i} \overline{e}_j\right).
\end{equation}
The clauses $\overline{\clause}_1,\ldots,\overline{\clause}_{\overline{m}}$ and $\overline{m}$ are related to $\clause_1,\ldots,\clause_m$ and $m$ in a complicated way.
Notice that ${\overline{\cal E}}$ is monotonic in terms of the negated variables $\overline{e}$,.
The probability that $\overline{e}_i = 1$ is $1-\wtx_i$, hence 
\begin{equation}
\Sat({\cal E},\vec{x}) = 1 - \Sat(\overline{{\cal E}}, 1-\vec{x})
\label{eq:duality}
\end{equation}
The complement ${\overline{\cal E}}$ induces a partial ordering on the power set of $\{1, \ldots , N\}$ that is the dual of the one induced by ${\cal E}$, as illustrated in \cref{fig:separatrix}.
This duality is at the heart of the Max Cut / Min Flow relationship \cite{Fulkerson2}.
To emphasize this duality, we refer to solutions of ${\cal E}$ as {\em struts} and solutions of $\overline{{\cal E}}$ as {\em cuts}.
The set of struts, as their name suggests, are the support of $\Sat({\cal E}, \vec{x})$;
the cuts, as we show below, are the usual cut sets in a graph induced by ${\cal E}$.

\subsection{Upper and lower bounds: parallel and series expressions}
\label{sec:bounds}
We refer to the conjunction ${\cal E}_s = \bigwedge_{i=1}^k e_i$ as the {\em series}  expression formed from these events, and the disjunction ${\cal E}_p = \bigvee_{i=1}^k e_i$ as the {\em parallel} expression. 
These expressions are duals of each other, with
\begin{equation}
\Sat({\cal E}_s,x) = x^k {\rm\ and\ } \Sat({\cal E}_p,x) = 1 - (1-x)^k 
\end{equation}
The identity Equation~\cref{eq:duality} is easily verified for this case.
Moreover, the series and parallel expressions bound the satisfiability of any non-trivial, finite, monotonic, homogeneous expression on exactly $N$ variables.
Considering nothing about ${\cal E}$ except that it contains $N$ distinct events, we still know there must be at least one solution, and the smallest probability it can have in the homogeneous case is $x^N$.
Applying the same logic to the dual expression and using \cref{eq:simpleBounds,eq:duality} gives
\begin{equation}
\Sat({\cal E}_s,x) = x^N \le \Sat({\cal E},x) \le 1-(1-x)^N = \Sat({\cal E}_p,x).
\label{eq:seriesParallel}
\end{equation}

\subsection{Implicit PSAT}
\label{sec:MC}
The perturbative approximations here are predicated on explicit expressions for {\em both} ${\cal E}$ and $\overline{{\cal E}}$ in terms of minimal struts and cuts, respectively.
Life rarely provides either, much less both, and almost never in terms of minimal clauses.
This section describes a simulation method for sampling the minimal clauses when they are unavailable {\em a priori}. 
This step adds another layer of approximation to the problem, and it is not yet clear how this affects the overall approximation error.
In principle, analyzing a small sample of minimal struts or cuts is equivalent to analyzing a simpler expression ${\cal E}$.

Although the individual events $e_i$ in a PSAT problem are not deterministic, deciding whether a particular set of events forms a solution is.
We will assume that ${\cal E}$ is embodied in a deterministic, binary oracle instead of an extensional disjunctive normal form expression.
A monotonic problem admits a continuous separatrix splitting the lattice of possible outcomes into cuts and struts,
as shown in \cref{fig:separatrix}.
Any path from the known cut at the top of the lattice to the known strut at the bottom must intersect the separatrix exactly once, so it can be located by a binary search in $O(\log N)$ steps.
However, not every configuration bordering the separatrix is minimal -- it is necessary to test for minimality, which could require up to $O(N)$ tests, i.e., calls to the oracle. 
If the configuration is not minimal, a minimal one can be found by walking along the separatrix, which requires at most an additional $O(N)$ steps.
In the worst case, it is possible that the process requires $O(N^2 \log N)$ calls to the oracle.
Average-case complexity is not obvious.
The search can be biased, for example towards or away from finding completely disjoint minimal sets or towards finding the smallest minimal sets.
Bespoke methods for specific classes of graphs or expressions, e.g., path-finding, may perform substantially better.

\section{Relation to network reliability}
\label{sec:graph}
Every monotonic expression can be mapped into a weighted, directed graph 
with two special vertices, $S$ and $T$, and every solution to the associated SAT problem corresponds to a path on that graph from $S$ to $T$.\footnote{
This construction can also be used to map any monotonic network reliability problem into an S-T version, demonstrating that S-T reliability is universal.
}
The partial order lattice itself is one such graph, where the vertices representing each clause and their descendants are identified and labeled $T$.
The mapping is not unique, but there is a single choice that is arguably the most natural representative, constructed as follows.
Beginning with the partial order lattice with vertices identified as above, recursively identify first, all the vertices with the same sets of incoming edges, and then, all the vertices with the same sets of outgoing edges.
A detailed algorithm is given in \cref{app:SAT2Graph}.
The satisfiability $\Sat$ is the probability that a random walker starting at $S$ will reach $T$ or, equivalently, the probability that a random subgraph of ${\cal G}$ constructed by choosing each edge with probability equal to its weight includes a path from $S$ to $T$.
This is the ``S-T'' or ``two-point'' network reliability introduced by Moore and Shannon\cite{Moore:56}.
The paths are the Feynman diagrams for the perturbative approximation to $\Sat$.
It is trivial to compute $\Sat$ on a tree, but loops, even acyclic ones, induce dependencies between different paths that are hard to deal with.

Notice that this construction can handle events that occur in both positive and negative senses as independent events, as long as we require that clauses that include both variables are ignored.
The resulting graphs can be split into two subgraphs that intersect only at $S$ and $T$.
Of course, every such event requires another split, so that, if there are $k$ such events, there will be $2^k$ separate subgraphs.
The methods described here are appropriate when the expression ${\cal E}$ is {\em mostly} monotonic, i.e., when $k \ll N$.
In principle, it is possible that more complicated constraints on the joint probability of two events could be handled, but that is beyond the scope of this work.
The inverse process -- constructing a SAT expression from a graph -- is exactly identifying minimal solutions.

For a monotonic system, $\Sat({\cal E}, \vec{x})$ is a monotonic polynomial with integer coefficients mapping the unit interval to itself.
Since each variable $e_i$ appears at most once in any clause or path, the reliability in the homogeneous case is a polynomial of degree at most $N$.
In the {\em thermodynamic limit}, $N \to \infty$, for many systems the partition function exhibits a discontinuity at a critical value of $x$, indicating a phase transition.
For finite $N$, there can be no discontinuity, but there can be a ``shadow'' of a discontinuity, i.e., an abrupt, nonlinear change in value over a small range of $x$, called the transition region.
This behavior limits the utility of Taylor expansions for $\Sat$, as indicated in the center panel of \cref{fig:example}.

\section{Perturbative methods}
\label{sec:methods}

\label{sec:Perturbative}
As discussed in \cref{sec:configs}, the reliability is not just the sum of total probabilities for each clause, because this overcounts the contribution of many configurations.
The Inclusion-Exclusion expansion correctly handles all the contributions, but only at the cost of increasing the number of terms to as many as $2^{|{\cal L}|}$.
Moreover, the terms form an alternating series with combinatorially large coefficients.
Applying the expansion to ${\cal E}$ (resp., $\overline{\cal E}$) produces a power series in $x$ (resp., $1-x$) -- i.e., a Taylor series at $x=0$ (resp., $x=1$) -- for the satisfiability.
These are the weak- and strong-coupling expansions of statistical physics.
Truncating either expansion at depth $\depth$ has the effect of truncating the associated Taylor series.
The truncated Taylor series do not respect unitarity, monotonicity, or duality, nor do they separately provide either an upper or lower bound on the answer.
However, combining the two truncated Taylor series using duality and imposing monotonicity on the result restricts the space of possible solutions and allows us to identify upper and lower bounds.

\subsection{Taylor series expansion(s)}
\label{sec:Taylor}

We can truncate the Inclusion-Exclusion expansion at any desired depth $\depth$ to obtain the first $\kappa(\depth)$ Taylor coefficients exactly, and thus an $O(x^{\kappa(\depth)})$ approximation, where $\kappa(\depth)$ is 1 less than the size of the smallest union of $\depth+1$ sets.
$\kappa(\depth)$ depends on ${\cal E}$, as illustrated in \cref{sec:example}:
\begin{equation}
\kappa(\depth) =  \min_{\depth+1-{\rm tuples\ } t}  \left|\bigcup_{i=1}^{\depth+1} \clause_{t_i}\right| - 1.
\label{eq:kappa}
\end{equation}
Minimality of the clauses guarantees that $\kappa(\depth)\ge \depth$.
In practice, it may be the case that $\kappa(\depth)\gg \depth$.

\subsection{Bezier polynomial representation}
\label{sec:Bezier}
Applying the Inclusion-Exclusion expansion to minimal configurations results in degree-$N$ polynomials in $x$ or $1-x$, i.e.,
\begin{align}
\Sat({\cal E}, x) = \sum_{k=0}^N \alpha_k x^k= \sum_{k=0}^N \overline{\alpha}_k (1-x)^k.
\label{eq:satTaylor}
\end{align}

The number of configurations with exactly $k$ events occurring is ${N \choose k}$ and the probability of each is $x^k(1-x)^{N-k}$.
Hence, Moore and Shannon \cite{Moore:56} suggested writing
\begin{equation}
\Sat({\cal E}, x) = \sum_{k=0}^N \beta_k {N \choose k} x^k (1-x)^{N-k}. 
\label{eq:Bezier}
\end{equation}
The transformation from $\vec{\alpha}$  to $\vec{\beta}$ is a change of basis in the vector space of polynomials of degree $N$ from the {\em power} basis to the {\em Bernstein} basis, whose  basis elements are: \cite{farouki2012bernstein,lorentz2013bernstein}
\begin{equation}
B(N, k, x) \equiv {N \choose k} x^k (1-x)^{N-k}.
\end{equation}
As summarized in the commutative diagram of \cref{fig:diagram}, a Taylor series can be thought of as a linear operator $\Tay$ from $ \mathbb{R}^{N+1}$ to polynomials of degree $N$ on $[0,1]$.
A Bezier polynomial is a (different) linear operator $\Bez$ from $ \mathbb{R}^{N+1}$ to polynomials of degree $N$ on $[0,1]$.
The representation in the Bernstein basis has many useful, well-known properties.
Here we will make use of the following:

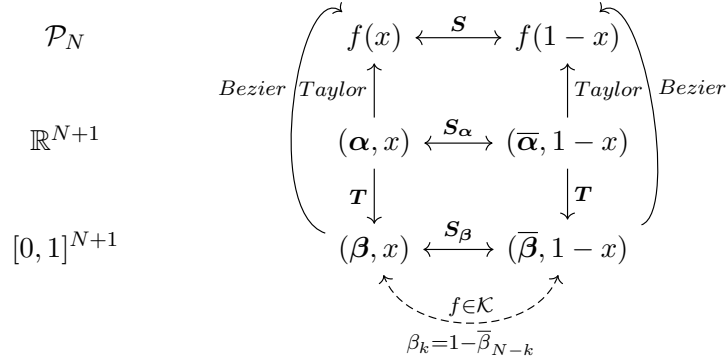
\begin{figure}[t]
\begin{center}
$\begin{tikzcd}
\Poly_N & \quad
& f(x) \arrow[r,leftrightarrow,"\vec{S}"]
& f(1-x) 
\\
 \mathbb{R}^{N+1} & \quad
& (\vec{\alpha},x) \arrow[r,leftrightarrow,"\vec{S_{\alpha}}"]
                                  \arrow[d,"\vec{T}"']
                                  \arrow[u,"Taylor"]
& (\vec{\overline{\alpha}},1-x) \arrow{d}{\vec{T}}
                                               \arrow[u,"Taylor"']
\\
\lbrack 0,1\rbrack^{N+1} & \quad
& (\vec{\beta},x)  \arrow[r,leftrightarrow,"\vec{S_{\beta}}"]
                                      \arrow[uu,controls={+(-1.5,0.5) and +(-1,0.8)},in=165,out=55,"Bezier"]
& (\vec{\overline{\beta}},1-x)
                                       \arrow[uu,controls={+(1.5,0.5) and +(1,0.8)},in=15,out=285,"Bezier"']
                                       \arrow[l,leftrightarrow,dashed,bend right,in=110,out=70,"f \in {\cal K}"'," \beta_k = 1-\overline{\beta}_{N-k}"]
\end{tikzcd}$
\caption{A commutative diagram illustrating relationships among: (1) $\vec{\alpha}$, Taylor coefficients at $x=0$; (2) $\vec{\overline{\alpha}}$, Taylor coefficients at $x=1$; (3) $\vec{\beta}$ and $\vec{\overline{\beta}}$, Bezier coefficients; and (4) the reflection $S$ which maps $x\leftrightarrow (1-x)$. The function $f(x)$ is a Taylor series in $x$ constructed from the coefficients $\vec{\alpha}$ or, equivalently, a Bezier polynomial in $x$ constructed from the coefficients $\vec{\beta}$. When $f$ is skew-symmetric about $x=\nicefrac{1}{2}$, e.g., $f(x) = 1 - f(1-x)$, then $\beta_k = 1-\overline{\beta}_{N-k}$, as indicated by the dashed curve. The resulting redundance in the Bezier coefficients can be used to bound the approximation error when not all Taylor coefficients are known.
}
\label{fig:diagram}
\end{center}
\end{figure}

\begin{itemize}
\item Bernstein basis functions $B(N,k,x)$ are strongly localized around the point $\nicefrac{k}{N}$. 
Hence they are kernel density estimators for functions on the unit interval. 

\item The Bernstein basis functions are invariant under the simultaneous operations $x\leftrightarrow 1-x$ and $k \leftrightarrow N-k$.
Hence, if $\vec{S_{\beta}}$ is a reflection, i.e.,
\begin{equation}
\vec{S_{\beta}}(\beta_0, \beta_1,\ldots,\beta_{N-1},\beta_N) \equiv (\beta_N, \beta_{N-1},\ldots,\beta_{1},\beta_0),
\label{eq:SDef}
\end{equation}
then
\begin{equation}
\Bez(\vec{\beta},x) = \Bez(\vec{S_{\beta}}(\vec{\beta},1-x)).
\label{eq:symmetry}
\end{equation}

\item The transformation from the power basis to the Bernstein basis is accomplished using a matrix closely related to Pascal's triangle.
Specifically, if
\begin{equation}
\vec{\beta} = \vec{T^{(N)}}\vec{\alpha},
\quad{\rm where\ } T^{(N)}_{k,j} \equiv  \left. {k \choose j} \right/{N \choose j} = \left.{N-j \choose k-j}\right/{N \choose k},
\label{eq:TDef}
\end{equation}
then $\Bez(\vec{\beta},x) = \Tay(\vec{\alpha},x)$.

\item The curve $(x,\Bez(\vec{\beta},x))$ is contained within the convex hull of the points $(\nicefrac{k}{N}, \beta_k)$. 
Hence if $\beta_0 = 0$, $\beta_N = 1$, and $0 \le \beta_k \le 1$ for all other $k$, then $\Bez(\vec{\beta},x)$ is also in the interval $[0,1]$. 
The analogous constraint on Taylor coefficients is not simple -- in fact it is most easily derived by way of $\vec{T}^{-1}$.

\item If the coefficients $\beta_k$ are monotonic in $k$, then $\Bez(\vec{\beta},x)$ is also monotonic. 
Once again, the analogous constraint on Taylor coefficients is not obvious.
Although the converse is not necessarily true in general, the semantics of the satisfiability's coefficients places a monotonicity constraint on them as shown in \cref{sec:interp}.

\item The derivative of a Bernstein polynomial at $x=0$ or $1$ is
\begin{equation}
\left.\frac{d^m}{dx^m} B(N,k,x)\right|_0 = 
\left\{\begin{array}{cl}
\frac{N!}{(N-m)!}(-1)^{m-k},& k \le m \le N\\
0& {\rm else}
\end{array}
\right.
\label{eq:BernDer}
\end{equation}

\item de Casteljau's algorithm for evaluating Bezier polynomials is numerically stable and does not require explicitly constructing the binomial coefficients. 
\end{itemize}

This basis was introduced by Bernstein to prove the Weierstrass approximation theorem \cite{farouki2012bernstein} by constructing the convergent single-parameter family of approximations 
\begin{equation}
f^{(N)} \equiv \sum_{k=0}^N f\left(\nicefrac{k}{N}\right) B(N,k,x).
\label{eq:BernPoly}
\end{equation}
Members of this family of approximants are referred to as {\em Bernstein polynomials} for $f$.
Here we instead construct an approximation whose first $m$ (resp., $\overline{m}$) derivatives 
at $x=0$ (resp., $x=1$) match those of $f$.
To emphasize the difference, we call these {\em Bezier polynomials}\cite{forrest1972interactive}.
They are also sometimes referred to as ``polynomials in Bernstein form''.

The symmetry of Bernstein basis functions shown in \cref{eq:symmetry} makes them ideal for representing the duality in PSAT problems.
Notice that if $\vec{\alpha}$ and $\vec{\overline \alpha}$ are, respectively, the Taylor coefficients of $f(x)$ and $1 - f(x)$ at $x=0$ and $1$,
then the symmetry $f(x) = 1 - f(1-x)$ allows us to write
\begin{subequations}
\begin{align}
\Tay(\vec{\alpha},x) &= 1 - \Tay(\vec{\alpha},1-x) = \Tay(\vec{\overline{\alpha}},1-x)  \\
\Bez(\vec{T}\vec{\alpha},x) &= \Bez(\vec{\overline{T}}\vec{\overline{\alpha}},1-x)\\
\Bez(\vec{\beta},x) &= \Bez(\vec{\overline{\beta}},1-x) = \Bez(\vec{S}\vec{\overline{\beta}},x).
\label{eq:reverse}
\end{align}
\end{subequations}
That is $\vec{\beta} = \vec{S}\vec{\overline{\beta}}$ or 
\begin{equation}
\beta_k = \overline{\beta}_{N-k},
\label{eq:reflect}
\end{equation}
as illustrated in the commutative diagram.

\subsection{Bounds and Interpolation}
\label{sec:interp}

The transformation matrix $\vec{T}$ is lower triangular.
Hence $\beta_k$ is completely determined by the values of $\alpha_j$ for $j \le k$.
From the first $\kappa$ coefficients of the Taylor series at $x=0$, we obtain the {\em first} $\kappa$ coefficients of the Bernstein representation; from the first $\overline{\kappa}$ coefficients of the Taylor series at $x=1$, we obtain the {\em last} $\overline{\kappa}$ Bernstein coefficients using Equation~\cref{eq:reverse}.
When $\kappa+\overline{\kappa}<N$, the 
coefficients $\beta_{\kappa},\ldots,\beta_{N-\overline{\kappa}}$ are undetermined.
Nonetheless, monotonicity allows us to place tight upper and lower bounds on $\beta_{k+1}$ given $\beta_k$. 
Although the convex hull property of Bezier polynomials ensures that monotonically increasing coefficients produce a monotonically increasing polynomial, the converse is not necessarily true.
The argument for the converse in this case provides insight into how $\vec{\beta}$ characterizes scale-dependent structure.

The number of solutions to the satisfiability problem in which exactly $k$ variables are true -- i.e., the number of struts in level $k$ of the partial order lattice --  is given by $n_k\equiv\beta_k {N \choose k}$ (which is thus the ``density of states'' function of statistical mechanics).
By monotonicity, each solution of size $k$ generates $N-k$ solutions of size $k+1$.
Of course, these solutions are not necessarily distinct.
Indeed, the coefficient $\beta_k$ encodes not only the number of minimal solutions of size $k$, but also 
how all the smaller minimal solutions overlap.
Since each vertex in level $k+1$ of the partial order lattice has only $k+1$ incoming edges,
\begin{equation}
\beta_{k+1} {N \choose k+1} = n_{k+1}\ge n_k \frac{N-k}{k+1} = \beta_k {N \choose k+1} \Longleftrightarrow \beta_{k+1} \ge \beta_k.
\end{equation}
On one hand, this bound is tight in the sense that there is an expression ${\cal E}'$ for which $\beta'_{k+1}=\beta'_k =\beta_{k}$;
on the other hand, it does not take advantage of all the known information about how minimal solutions overlap that is contained in $\{\beta_0,\ldots,\beta_{k-1}\}$.

An upper bound can be obtained from the lower bound of the dual problem.
Together, these constrain $\beta_{\kappa} \le \beta_k \le \beta_{N-\overline{\kappa}}$ for 
$\kappa < k < N-\overline{\kappa}$.
These bounds are shown in the center panel of \cref{fig:example}.

There are several intuitively appealing interpolants between the bounds, including:
\begin{enumerate}
\item linear interpolation, 
$\widehat{\beta}_k = \beta_{\kappa} + \frac{k-\kappa}{N-\overline{\kappa}-\kappa} (\beta_{N-\overline{\kappa}}-\beta_{\kappa})$;
\item logarithmic interpolation,
$\ln \widehat{\beta}_k = \ln\beta_{\kappa} + (\frac{k-\kappa}{N-\overline{\kappa}-\kappa})(\ln\beta_{N-\overline{\kappa}}-\ln \beta_{\kappa})$;
\item the expected value of $n_k$ under an assumption that the graph is ``structureless'' at these scales -- for example, lacking any minimal solutions of sizes $\kappa < k < N-\overline{\kappa}$, and whose minimal solutions outside that range overlap randomly. See \cref{sec:expected} for more details.
\end{enumerate}

\subsection{Hybrid estimation}
\label{sec:hybrid}
The upper and lower bounds developed in \cref{sec:bounds} define feasible regions that are narrowest near $x=0$ and $1$.
Details of the region near $0<x<1$ are determined by Bezier coefficients $\beta_k$ for $k \sim xN$.
Estimates of $\beta_k$ from any source can dramatically narrow the uncertainty in $\Sat$, although estimates alone do not change the bounds.
One such estimate can be provided by Monte Carlo simulation.
By definition, a fraction $\beta_k$ of the subsets of exactly $k$ events satisfies ${\cal E}$.
Hence, we can estimate $\beta_k$ for any $k$ to any desired confidence by evaluating ${\cal E}$ on a random sample of subsets of events.
Another estimate, for the special point $\wtx=\nicefrac{1}{2}$ where each event is equally likely to occur or not, is provided by $2^{-N}{\cal N}$, where ${\cal N}$ is the number of solutions of the corresponding deterministic satisfiability problem.

\subsection{Sensitivity analysis}
\label{sec:sens}

The satisfiability $\Sat({\cal E},\vec{x})$  is a multi-affine function of the event probabilities, since multiple appearances of the same event within a single conjunction can be reduced to a single occurrence.
That is, it can be written as $\prod_{i=1}^N (a_i + b_i \wtx_i)$.
Maximizing or minimizing the satisfiability is thus, in principle, not difficult once the coefficients have been determined.
However, a common problem is to optimize satisfiability under correlated constraints on the event probabilities, such as constraining them to lie in a subspace of $\Re^N$.

Restricted to a subspace of dimension $M$, the satisfiability can be thought of as a smooth (because it is a multinomial) $M$-dimensional manifold.
An approximation allows us to apply standard tools such as sensitivity analysis or differential geometry to this manifold. 
For example, suppose the probability of each event depends linearly on a finite resource such as energy, bandwidth, vaccine, or human time and effort.
Re-allocating  resources from one set of events to another can be modeled as a perturbation in the corresponding direction.
The partial derivatives $\frac{\partial}{\partial x_i} \Sat({\cal E},\vec{x})$ are a differential form of the leave-one-out or Birnbaum importance \cite{birnbaum69} of event $E_i$.
In the graphical representation of the problem, they define a notion of graph derivatives.

\section{Homogeneous Example}
\label{sec:example}

The following simple, analytically tractable example illustrates an important point: the relative contribution of different events to the overall satisfiability depends on the probability of the individual events, {\em even in a homogeneous system}.
Concretely, think of two strains of an infectious disease with different transmissibilities spreading over a human contact network in which all contacts are identical. 
Obviously, the more transmissible strain is more likely to infect any given person;
less obviously, the particular contacts whose removal most reduces the probability of infecting that person may differ.

\begin{figure}[htbp]
\begin{center}
\includegraphics[width=0.35\textwidth,trim= 24 -48 -24 24]{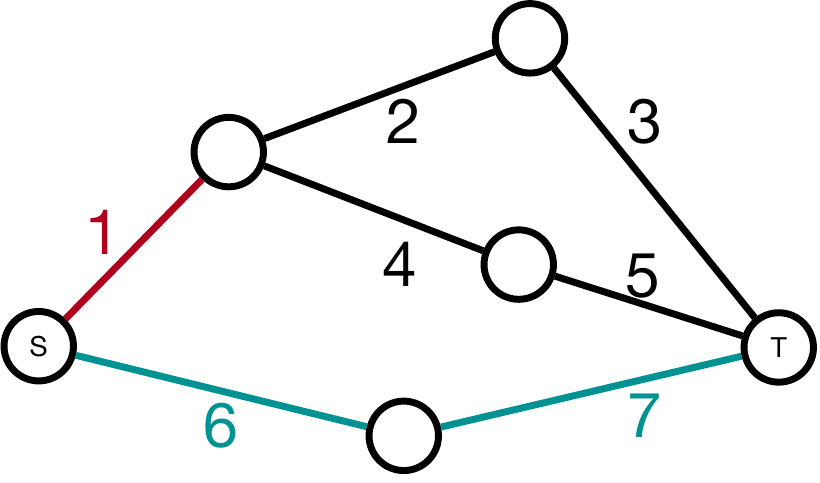}
\hfil
\includegraphics[height = 0.35\textwidth]{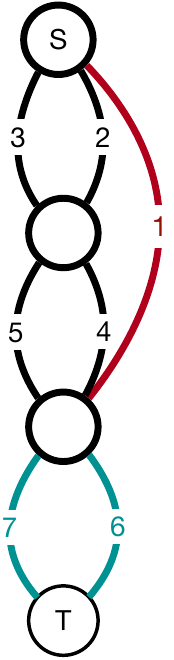}
\caption{Graphical representations of the expression ${\cal E}$ in Equation~\cref{eq:exampleE} (left panel) and its dual (right panel).
Edges with the same contribution to the satisfiability, as evident by symmetry, are colored the same.}
\label{fig:toyGraph}
\end{center}
\end{figure}

Consider the set of $N=7$ events and the expression
\begin{equation}
{\cal E} = (e_1 \land e_2 \land e_3 ) \lor (e_1 \land e_4 \land e_5) \lor (e_6 \land e_7).
\label{eq:exampleE}
\end{equation}
The clauses (and hence the minimal struts) are defined by the sets $\clause_1 = \{1,2,3\}$; $\clause_2=\{1,4,5\}$; and $\clause_3=\{6,7\}$.
An equivalent network reliability problem is the probability of reaching $T$ from $S$ on the graph in the left panel of \cref{fig:toyGraph}, constructed using the algorithm in \cref{app:SAT2Graph}.
The three minimal struts 
are the simple paths from $S$ to $T$ in this graph;
the 10 minimal cuts are easily seen to be those given in \cref{tab:cuts}, which are $(S,T)$ cut sets for this graph.
The right panel of \cref{fig:toyGraph} shows a graph for the dual expression.
The minimal struts and cuts for the graph in the left panel are the minimal cuts and struts, respectively, for the graph in the right panel.

\begin{table}[htp]
\caption{The ten minimal cuts for the example expression in Equation~\cref{eq:exampleE}.}
\begin{center}
\begin{tabular}{ccccc}
$\{1,6\}$ & $\{2,4,6\}$ & $\{2,5,6\}$ & $\{3,4,6\}$ & $\{3,5,6\}$ \\
$\{1,7\}$  & $\{2,4,7\}$ & $\{2,5,7\}$ & $\{3,4,7\}$ & $\{3,5,7\}$
\end{tabular}
\end{center}
\label{tab:cuts}
\end{table}%

\subsection{Satisfiability}

Consider the first clause, $e_1 \land e_2 \land e_3$.
The probability of the minimal strut corresponding to this clause is $p(\{1,2,3\}) = x^3 (1-x)^4$;
its total probability (including all its descendants)  is $p_t(\{1,2,3\}) = x^3$.

Because there are only three minimal struts, the Inclusion-Exclusion expansion for $\Sat({\cal E}, x)$ contains only $2^3-1$ terms and can easily be written down by inspection:
\begin{multline}
\Sat({\cal E}) = p(e_1 \land e_2 \land e_3) + p(e_1 \land e_4 \land e_5) + p(e_6 \land e_7)\nonumber\\
- p(e_1 \land e_2 \land e_3 \land e_4 \land e_5) - p(e_1 \land e_2 \land e_3 \land e_6 \land e_7) - p(e_1 \land e_4 \land e_5 \land e_6 \land e_7) \nonumber\\
+ p(e_1 \land e_2 \land e_3 \land e_4 \land e_5 \land e_6 \land e_7).
\end{multline}
Substituting in the total probability for each term yields the power series in $x$ (i.e., the Taylor series at $x=0$):
\begin{equation}
\Sat({\cal E}, x) = x^2 + 2x^3 - 3x^5 + x^7.
\end{equation}
There are $2^{10}-1=1023$ terms in the Inclusion-Exclusion expansion in terms of minimal cuts, so it is more difficult to write down,\footnote{
Obviously, since this is more than the number of possible distinct configurations, the expansion can be greatly simplified.
}
 but it yields the following power series in $y=1-x$ (i.e., the Taylor series at $x=1$):
\begin{equation}
\Sat({\cal E}, 1-x) = 1 - \overline{\Sat}({\cal E}, y) = 1-2y^2-7y^3+20y^4-18y^5+7y^6-y^7 
\end{equation}
Equivalently,
\begin{subequations}
\begin{align}
\vec{\alpha} &= (0,0,1,2,0,-3,0,1)
\label{eq:exampleCoefs}
\\
\vec{\overline{\alpha}} &= (1,0,-2,-7,20,-18,7,-1).
\label{eq:exampleBarCoefs}
\end{align}
\end{subequations}
Using Equation~\cref{eq:TDef}, we find for this expression
\begin{subequations}
\begin{align}
\vec{\beta} &= (0, 0, \nicefrac{1}{21}, \nicefrac{1}{5}, \nicefrac{18}{35}, \nicefrac{19}{21}, 1, 1)
\label{eq:exampleBeta}
 \\
\vec{\overline{\beta}} &= (1,1,\nicefrac{19}{21}, \nicefrac{18}{35}, \nicefrac{1}{5}, \nicefrac{1}{21}, 0, 0).
\end{align}
\end{subequations}
Notice that, although the Taylor coefficients are not related in any obvious way, the Bernstein coefficients satisfy Equation~\cref{eq:symmetry}.

\begin{table}[htp]
\caption{Degree of largest exact term in an Inclusion-Exclusion expansion of the example expression \cref{eq:exampleE} truncated at depth $\depth$.}
\begin{center}
\begin{tabular}{r|cccccccccc}
$\kappa(d)$ & 4 & 6 & 7 
\\
$d$ & 1 & 2 & 3 & 4 & 5 & 6 & 7 & 8 & 9 & 10
\\
$\overline{\kappa}(d)$ & 2 & 4 & 4 & 5 & 5 & 5 & 5 & 6 & 6 & 7
\end{tabular}

\end{center}
\label{tab:kappa}
\end{table}%

The functions $\kappa(d)$ and $\overline{\kappa}(d)$ defined in \cref{eq:kappa} for this expression are tabulated in \cref{tab:kappa}.
Now suppose we truncate the Inclusion-Exclusion expansions at depth $\depth=1$, i.e., not even considering pairs of minimal struts or cuts.
The resulting Taylor expansion at $x=0$ is exact up to and including terms of order $x^{\kappa(1)}=x^4$, while the Taylor expansion at $x=1$ is exact up to and including terms of order $(1-x)^{\overline{\kappa}(1)}=(1-x)^2$.
These estimates are shown in the center panel of \cref{fig:example}. 
Although each approximation tracks the exact function poorly through the transition region,
the Bernstein representation produced by combining the two Taylor expansions determines the value of every coefficient -- and thus, the function itself -- exactly, since $\kappa(1)+ \overline{\kappa}(1) = N-1$.
Suppose, for the sake of argument, that $\kappa(1)$ were smaller, say 3.
Monotonicity and linear interpolation between known coefficients define three possible values for $\beta_4$: a lower bound of $\nicefrac{1}{5}$, an upper bound of $\nicefrac{19}{21}$, and an interpolated value of $\nicefrac{58}{105}$. 
The resulting bounds and interpolation are shown in the right panel of \cref{fig:example}, along with the exact result.

\section{Heterogeneous Systems}
\label{sec:heterogeneous}
Suppose we are given a problem instance, i.e., an expression ${\cal E}$ with specified event probabilities $\vec{\wtx}$. 
Parameterized forms of the probabilities are often, but not always, part of the problem specification.
For example, the events $\vec{E}$ may be generated by a Poisson process operating for a time $\eparam$ or, equivalently, by interactions with a coupling constant $\eparam$.
The transmission of infectious disease from one host to another is often modeled as a Poisson process whose probability depends on the overall transmissibility of the pathogen, $\eparam$, and the duration of contact between the hosts, $\rate$.
More generally, any dynamical system whose configurations are probabilistically distributed (in time or across ensembles of identically prepared systems) as an exponential of a property of the configuration can be thought of as a collection of Poisson events with heterogeneous rates.\footnote{
This does not imply that the dynamics of such systems are Poisson processes, only that steady-state or equilibrium distributions can be modeled using them.
}
Such systems include statistical mechanical systems governed by Boltzmann distributions and field-theoretical systems governed by a least-action principle.

Applying the duality symmetry to the weak- and strong-coupling perturbation series relied on extending a particular problem instance into a one-parameter family whose solution smoothly interpolated between $0$ and $1$.
This parameter completes the transition from satisfiability as a binary function of binary variables to a continuous function of a continuous variable.
For the heterogeneous problem, we proceed analogously, first constructing a mapping to a two-parameter representation in which the two degrees of freedom contained in $\wtx_1$ and $\wtx_N$ are replaced with a ``location'' $\loc$ for a homogeneous case and an envelope $\wid$ capturing the extent of heterogeneity.
This mapping is convenient for distinguishing the effects of stronger interactions from those of heterogeneity through standard sensitivity analyses.
Then, in order to apply perturbative methods, we consider curves in this two-dimensional space along which we can ensure the satisfiability is monotonic.

Generally, the probabilities $\wtx_i$ are not unrelated, even when they are not identical.
That is, we can consider them all as functions $x_i(\param)$, of a single parameter $\param\in[0,1]$ with the special values $x_i(0)=0$, $x_i(1)=1$, and $x_i(\param^*)=\wtx_i$, for some $\param^*$ independent of $i$.
Colloquially, we may say that $E_1$ is ``twice as likely as'' $E_2$, for example.
Such a statement does not specify a unitary parameterization over the whole domain, since it cannot be true for $x_2 > \nicefrac{1}{2}$. 

If the parameterization is not specified, we are free to create an arbitrary one.
In \cref{sec:polyParam}, we describe a parameterization that is applicable in the absence of a parametric form for $\wtx_i$;
in \cref{sec:logParam}, a different parameterization for a specific, important case when the events are generated by  Poisson processes with different rates.

\begin{figure}[tbp]
\begin{center}
\includegraphics[width=0.45\textwidth]{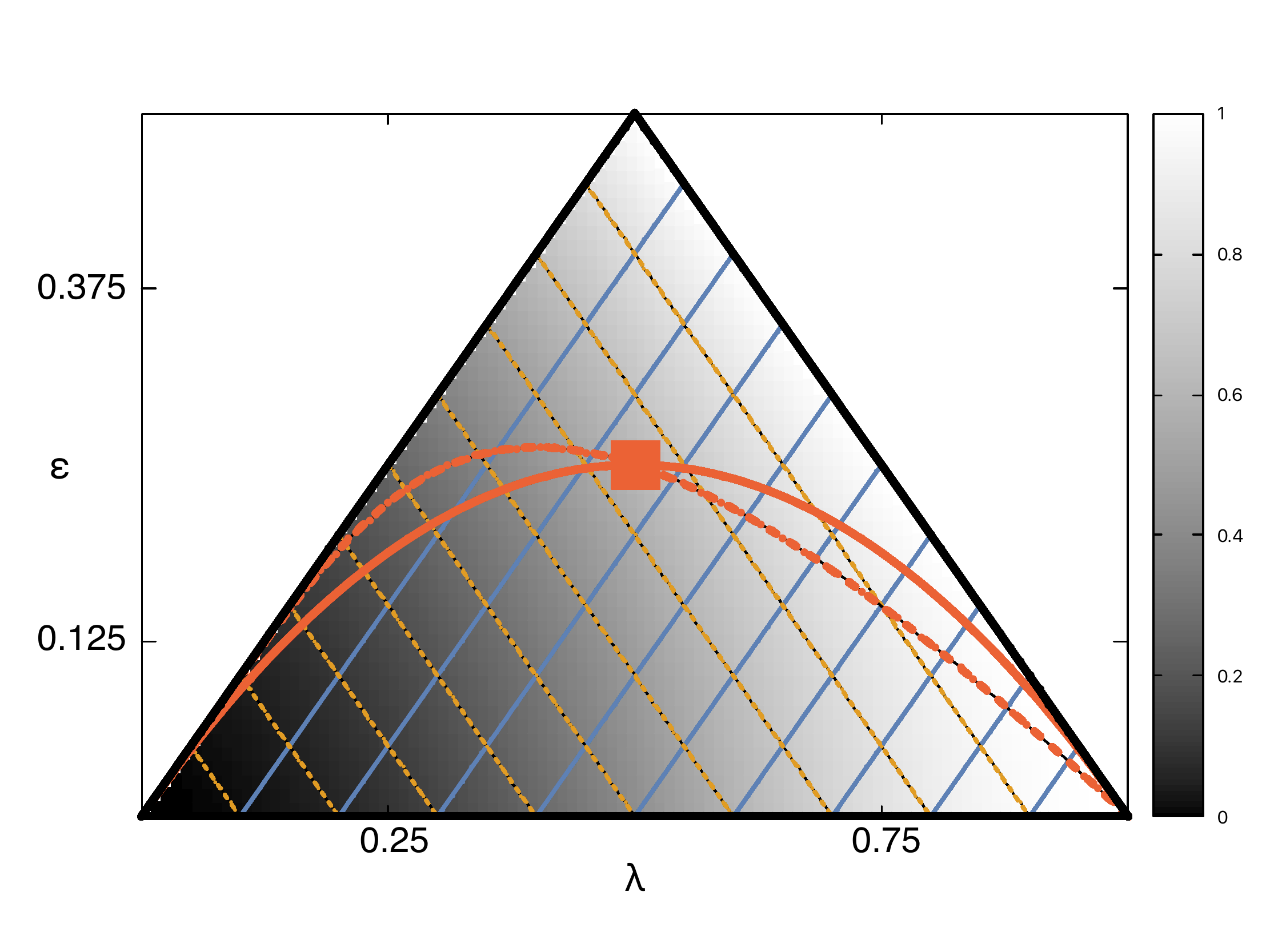}\hfil
\includegraphics[width=0.35\textwidth, trim= 24 -64 -24 24]{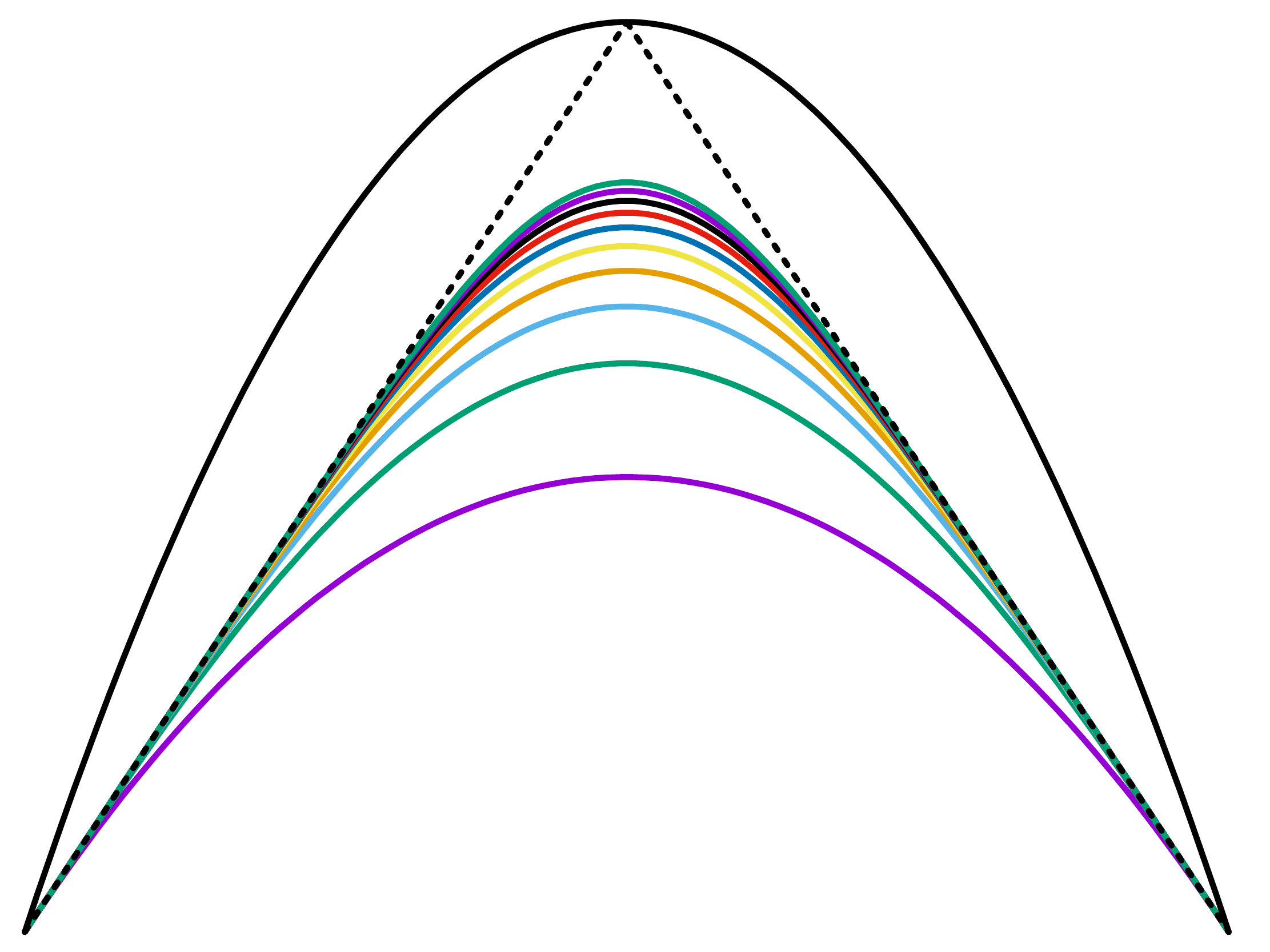}
\caption{
(Left) 
The coordinate mapping from $\wtx_1,\wtx_N$ to $\loc,\wid$ defined by \cref{eq:hetero}. 
Lines of constant $\wtx_1$ are mapped into blue solid curves;
lines of constant $\wtx_N$ are mapped into yellow dashed curves.
The background is shaded by the satisfiability for the example expression when $\wtx_B$ is the mean of $\wtx_A$ and $\wtx_C$.
The red curve is the parametric curve $(\loc(\param),\wid(\param))$ defined under the polynomial decomposition of \cref{sec:polyParam} for the case $\wtx_1=\nicefrac{1}{4}$, $\wtx_N=\nicefrac{3}{4}$. 
The square on the curve is the location of the example case. 
Notice that if $\wid$ were any larger, the parameterized curve would violate the unitarity constraints $\wid \le \min(\loc,1-\loc)$.
The dashed red curve is the parametric curve defined under the logarithmic decomposition of \cref{sec:logParam} for this case. 
(Right) The first 10 Bezier polynomial approximants defined in \cref{eq:BernPoly} to the unit triangle are shown inscribed in the triangle. 
The quadratic that passes through the points $(0,0)$, $(\nicefrac{1}{2},1)$, and $(1,0)$ is shown in black for comparison.
}
\label{fig:mappings}
\end{center}
\end{figure}

\subsection{Polynomial decomposition}
\label{sec:polyParam}

A useful polynomial decomposition is illustrated in \cref{fig:mappings}:
\begin{subequations}
\label{eq:hetero}
\begin{align}
x_i = \loc + \weight_i  \wid &{\rm\ and\ } \weight_i = (\wtx_i - \wtloc) / \wtwid,{\rm\ where\ }\\
 \wtloc \equiv \nicefrac{(\wtx_1+\wtx_N)}{2} &{\rm\ and\ } \wtwid \equiv \nicefrac{(\wtx_N-\wtx_1)}{2}.
\end{align}
\end{subequations}

This mapping results in a particularly simple transformation between the original problem and its dual, for which $\overline{\wtx}= 1 - \wtx$ and the ordering of events is reversed (i.e., $\overline{\wtx}_N=1-\wtx_1$):
\begin{align}
\overline{\loc} =1 - \loc;\quad
\overline{\wid} &= \wid;\quad \overline{\weight}_i =  -\weight_{N+1-i}.
\label{eq:dualParams}
\end{align}

Repeating the analysis of \cref{sec:Bezier} 
shows that the satisfiability can be written 
as a finite Taylor series in two variables analogous to \cref{eq:satTaylor}:
\begin{subequations}
\begin{equation}
\Sat({\cal E},\loc,\wid,\vec{\weight}) = \sum_{k=0}^N\sum_{\ell=0}^{N-k} \alpha_{k,\ell} \loc^{\ell} \, \wid^{k}
\label{eq:heteroTaylor}
\end{equation}
or, in a mixed Bernstein-Taylor basis: 
\begin{align}
\Sat({\cal E},\loc,\wid,\vec{\weight})  
&= \sum_{k=0}^{N}  \sum_{\ell=0}^{N-k}\gamma_{k,\ell}(\vec{\weight}) B(N-k,\ell,\loc) \wid^{k}.
\label[equation]{eq:gamma}
\end{align}
\end{subequations}
Furthermore, the gradient of the satisfiability at any point in the $\loc-\wid$ plane is easily computed in terms of $\vec{\gamma}$:
\begin{subequations}
\begin{align}
\frac{\partial\Sat({\cal E},\loc,\wid)}{\partial\loc} =& \sum_{k=0}^{N} \sum_{\ell=0}^{N-k-1}   (\gamma_{k+1,\ell} - \gamma_{k,\ell})B(N-k-\ell-1,\ell,\loc) \\
\frac{\partial\Sat({\cal E},\loc,\wid)}{\partial\wid} =& \sum_{k=1}^{N} \sum_{\ell=0}^{N-k-1}   k \gamma_{k,\ell} B(N-k-\ell,\ell,\loc)  \wid^{k-1}.          
\end{align}
\end{subequations}

Duality provides a generalization of the reflection symmetry previously obtained for homogeneous systems \cref{eq:reflect}:
\begin{equation}
\overline{\gamma}_{k,\ell}(-\vec{\weight})  = \gamma_{k,N-k-\ell}(\vec{\weight}) 
\label{eq:heteroReflect}
\end{equation}
Hence, order by order in $\wid$, we can use methods of \cref{sec:dual} to evaluate both the lowest and highest order coefficients in $\loc$.
Interpolation is not as easy, because $\Sat$ is not necessarily monotonic order by order in $\wid$.
Instead, we introduce parametric curves in the $\loc-\wid$ plane.
As long as $x_i(\param)$ is monotonic, the satisfiability must also be monotonic in $\param$, so we can apply the interpolation scheme developed for the homogeneous case in \cref{sec:interp} to $\Sat({\cal E},\param)$ to develop bounds on the satisfiability that are exact up to some order in $\param$.

To simplify the perturbative analysis, we choose $\loc(\param) = \param$, which implies $\param^* = \loc(\param^*) = \wtloc$, and normalize the weights $\weight_1 = -\weight_N=1$, which implies $\wid(\param^*) = \wtwid$.
Then we design a function $\wid(\param)$  that 
\begin{enumerate}
\item  satisfies unitarity: $\forall \param \in [0,1], \, 0 \le \wid(\param) \le \min(\param,1-\param) \le \nicefrac{1}{2}$;
\item  remains invariant under duality: $\param\leftrightarrow 1-\param$;
\item maintains the semantics of $\loc$ and $\wid$ as the midpoint and half-width of $[x_1(\param),x_N(\param)]$;
\end{enumerate}

Along the parameterized curve, \cref{eq:heteroTaylor} becomes
\begin{equation}
\Sat({\cal E},\param,\vec{\weight}) = \sum_{k=0}^N\sum_{\ell=0}^{N-k} \alpha_{k,\ell} \loc(\param)^{\ell} \, \wid(\param)^{k}.
\end{equation}
Because of the first condition above, $\wid(\param)$ vanishes at the points $t=0$ and $1$ where the exact solution to the homogeneous problem is known, and the Taylor expansion can be readily determined at both points.

An appealing choice for $\wid(\param)$ is a low-degree polynomial.
When $\wid(\param)$ is a degree-$m$ polynomial, the multinomial Taylor series reduces to a degree-$mN$ polynomial in $\param$.
However, it may require extremely large $m$ to maintain unitarity while simultaneously representing the full range of probabilities $\wtx_N - \wtx_1$.
For example, as illustrated in the left panel of \cref{fig:mappings}, a quadratic $\wid(\param)$ meeting these constraints cannot represent problems in which $\wtx_N-\wtx_1>\nicefrac{1}{2}$.
The Bernstein approximants defined in \cref{eq:BernPoly} for the unit triangular function, $T(\param) = 1 - 2|\param-\nicefrac{1}{2}|$, provide a convenient way of describing the lowest-order polynomial that can represent the full range of probabilities.
That is,
\begin{align}
\wid_{2n}(\param) &= \sum_{k=0}^{2n} T(\nicefrac{k}{n} )B(n,k,\param) 
= 1- \frac{1}{n} \sum_{j=-n}^n |j| \left[B(2n,n+j,t)+B(2n,n-j,t)\right].
\end{align}
where $n$ is the smallest integer such that $\wtx_N-\wtx_1 \le 2 \wid(\nicefrac{1}{2}) = 1-2^{1-2n}$.
For $n=1$, this reduces to $\wid_2(\param) = 2\param(1-\param)$, and, as noted above, $\wtx_N - \wtx_1$ must be less than or equal to $\nicefrac{1}{2}$.
The functions $\wid_{2n}$ for $n\le10$ are illustrated in the right panel of \cref{fig:mappings}.
Substituting $x_i(\param) = \param + w_i\wid_2(\param)$ into the Taylor series \cref{eq:heteroTaylor} gives
\begin{align}
\Sat({\cal E},t) &= \sum_{k=0}^{3N} \alpha'_k \param^k,{\rm\ where\ }
\alpha'_k = \sum_{u=0}^N \sum_{r=0}^{N-u} 2^r (-1)^{k-u-r}  {k\choose k-u-r} \alpha_{r,u}.
\end{align}

\subsection{Logarithmic decomposition}
\label{sec:logParam}

When parametric forms of the probabilities $\vec{x}$ are specified, it is more natural to use them than to pick an arbitrary parameterization as in \cref{sec:polyParam}.
Here we develop a decomposition and related parameterization appropriate to the important case of independent Poisson processes with heterogeneous rates, i.e., $x_i(\eparam) = 1 - e^{-\rate_i\eparam}$.
In this case, it is more useful to perform the decomposition in a logarithmic space, using the geometric, instead of the arithmetic, mean.
Choosing $\eparam^*=1$, the analogues of \cref{eq:hetero} become:
\begin{subequations}
\label{eq:heteroExp}
\begin{align}
\wtx_i(\eparam) = 1-e^{-(\rate+\weight_i\ratediff)\eparam}&{\rm\ and\ } \weight_i = -\left[\ln(1-\wtx_i)-\rate\right]/\ratediff,{\rm\ where\ }\\
\rate \equiv\nicefrac{(\rate_1+\rate_N)}{2} = -\ln\sqrt{(1-\wtx_1)(1-\wtx_N)}&{\rm\ and\ }
\ratediff\equiv\nicefrac{(\rate_1-\rate_N)}{2} = \ln\sqrt{\frac{1-\wtx_N}{1-\wtx_1}}.
\end{align}
\end{subequations}
When the rates $\rate_i$ are all rationally related -- as is common -- the exponent can be written as the ratio $\nicefrac{\rate_i}{\rate_N}=\nicefrac{\numer_i}{\denom}$ where $\denom$ is the greatest common divisor of  $\rate_i$ and $m_i$ is an integer greater than or equal to 1.
Then rescaling the parameter gives
\begin{align}
x_i(\param) = 1 - t^{m_i}, {\rm\ where\ } \param(\eparam) = e^{-\nicefrac{\rate_1\eparam}{\denom}}.
\end{align}
The satisfiability can thus be written as a degree-$M$ polynomial in $t$, where $M = \sum_i m_i$.

\section{Heterogeneous example}
\label{sec:example2}
Consider the same example as in \cref{sec:example}, except with heterogeneous probabilities $\vec{\wtx}$.\footnote{
Code for reproducing these results in Mathematica is available from the authors.
}
Suppose events in each equivalence class have the same probability:
\begin{equation}
\wtx_i = \left\{
\begin{array}{lr}
x_A = 1-y_A & i = 1\\
x_B=1-y_B & 2 \le i \le 5\\
x_C=1-y_C & 6 \le i \le 7.
\end{array}
\right.
\end{equation}
Then a straightforward analysis of the three minimal struts gives
\begin{align}
\Sat({\cal E},\vec{x}) 
&= x_C^2 + (1 - x_C^2) x_A x_B^2(2 - x_B^2)
\end{align}
and, from the 10 minimal cuts:
\begin{align}
\Sat({\cal E},\vec{x}) &=1- y_C(2-y_C) \left[y_A + (1-y_A)y_B^2(2-y_B)^2\right]
\end{align}
In this simple example, these polynomials in three variables are amenable to direct analysis.
Even here, though, if all the probabilities $\wtx_i$ were distinct, $\Sat$ and $\overline{\Sat}$ would be multinomials in seven dimensions, with potentially as many as $128$ terms.

As discussed in \cref{sec:sens}, the derivative of $\Sat(\wtx_B+2\alpha,\wtx_B, \wtx_B-\alpha)$ evaluated at $\alpha=0$ gives the effects of redistributing resources among the interactions that enable the single event in class $A$ and the two events in class $C$.
For this example, we find
\begin{equation}
\left.\frac{d\Sat(\wtx_B+\alpha,\wtx_B, \wtx_B-2\alpha)}{d \alpha} \right|_{\alpha=0} = -2\wtx_B \left(\wtx_B^3-2 \wtx_B+1\right).
\label{eq:derTheta}
\end{equation}
This function changes sign at $\wtx_B\approx0.61$, signifying that the relative importance of events in class $A$ and $C$ depends on the probability of the events in class $B$.

\begin{figure}[tbp]
\begin{center}
\includegraphics[width=0.65\textwidth]{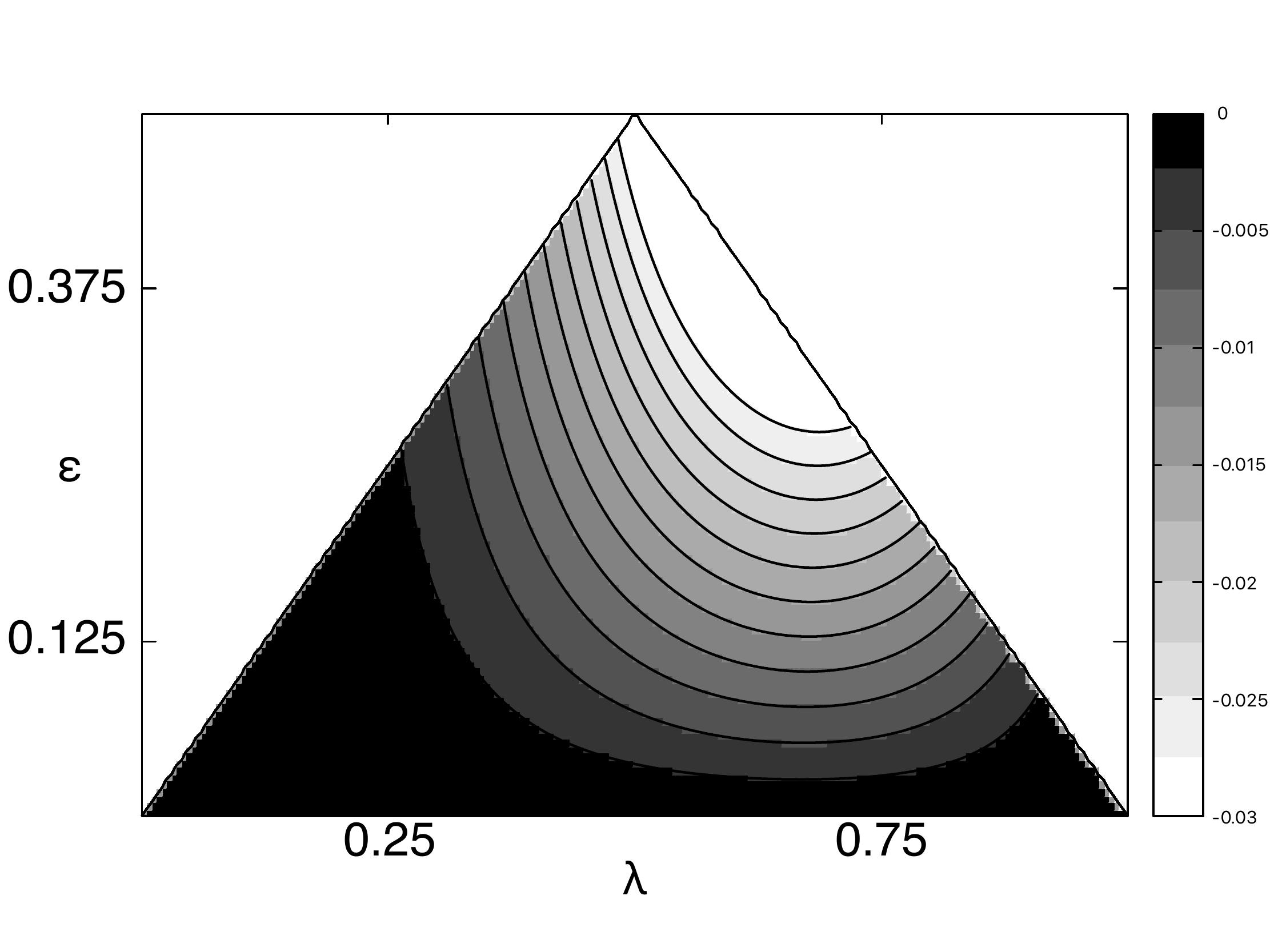}
\caption{
Approximation error in the satisfiability $\Sat$ for the heterogeneous example of \cref{sec:example2} with $(\wtx_A,\wtx_B,\wtx_C) = \nicefrac{1}{4},\nicefrac{3}{8},\nicefrac{1}{2}$, using the decomposition described in \cref{sec:polyParam} and a linear interpolation between the Bezier coefficients determined by a depth-2 expansion at $\loc=0$ and a depth-1 expansion at $\loc=1$. 
}
\label{fig:approxError}
\end{center}
\end{figure}

\subsection{Transforming to $\loc-\wid$ space}
\begin{figure}[tbp]
\begin{center}
\includegraphics[width=0.475\textwidth]{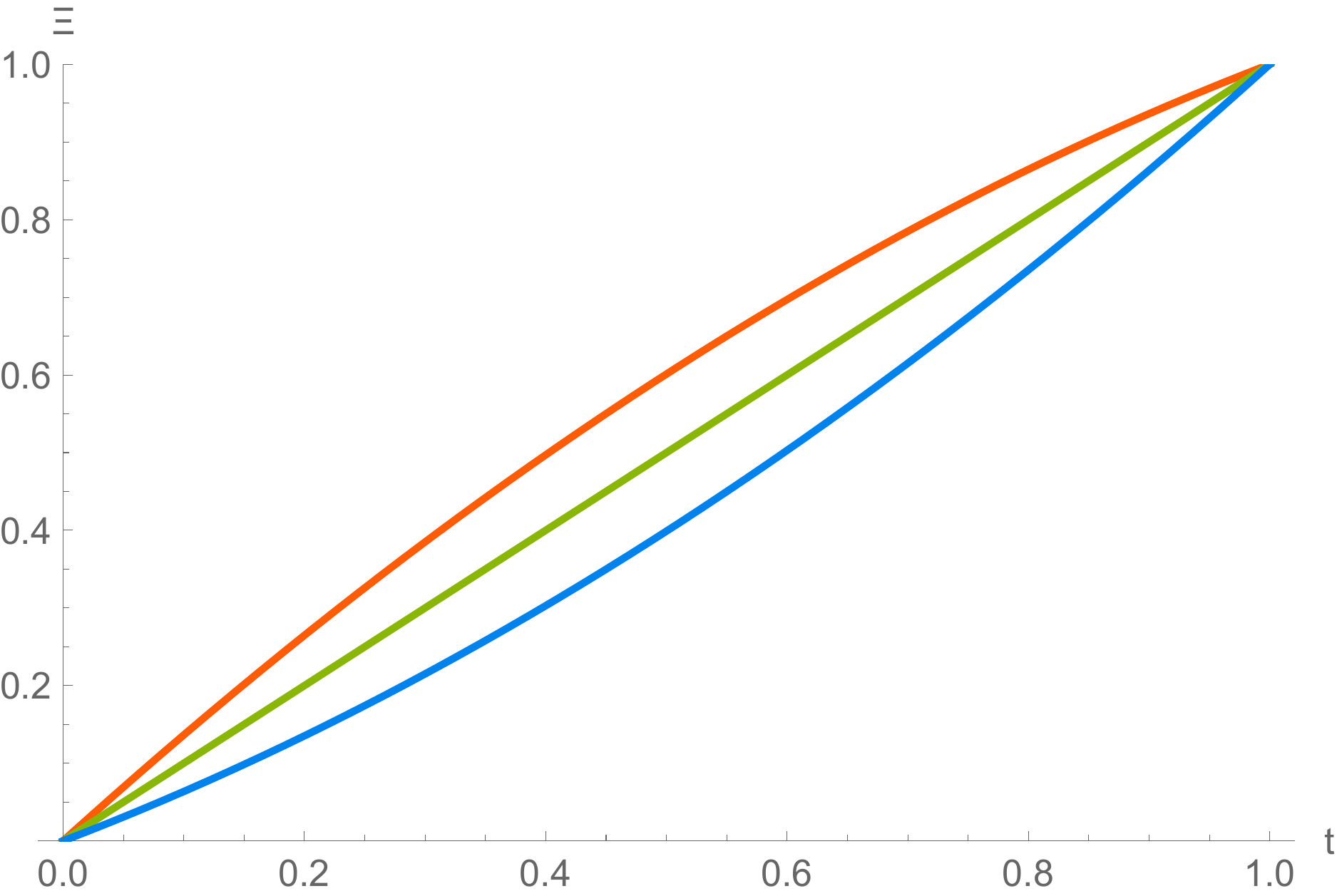}\hfil
\includegraphics[width=0.475\textwidth]{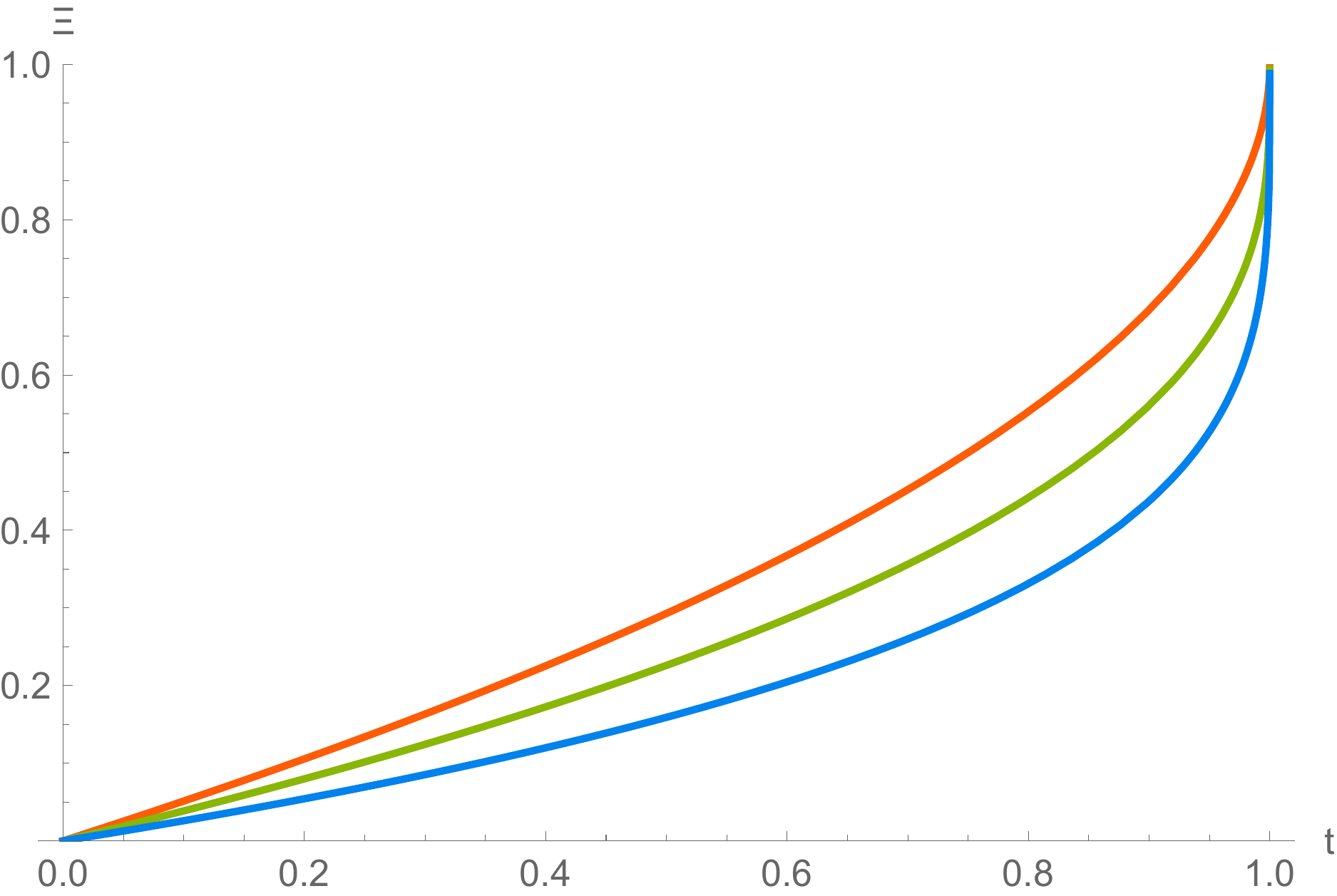}\\
\includegraphics[width=0.475\textwidth]{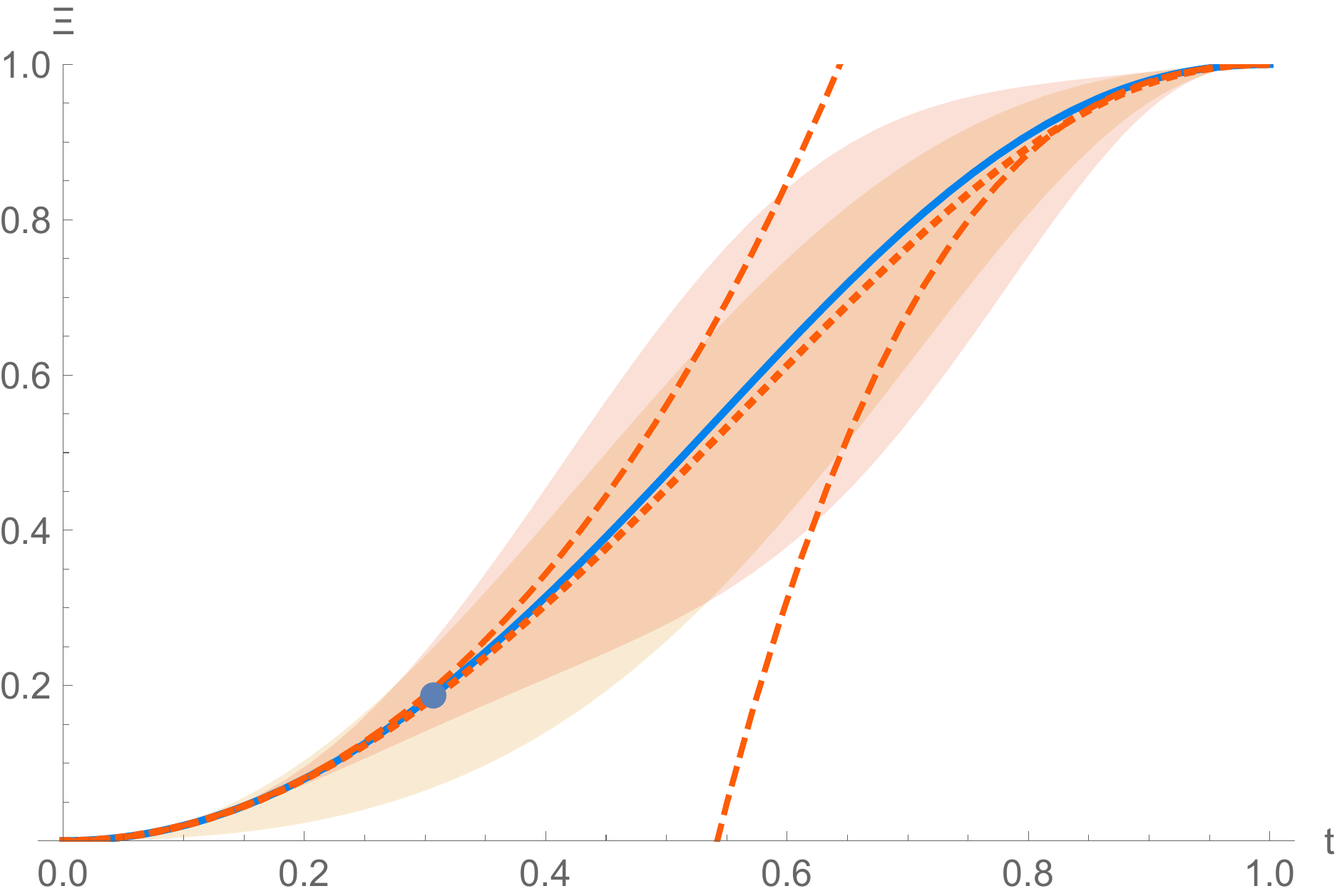}\hfil
\includegraphics[width=0.475\textwidth]{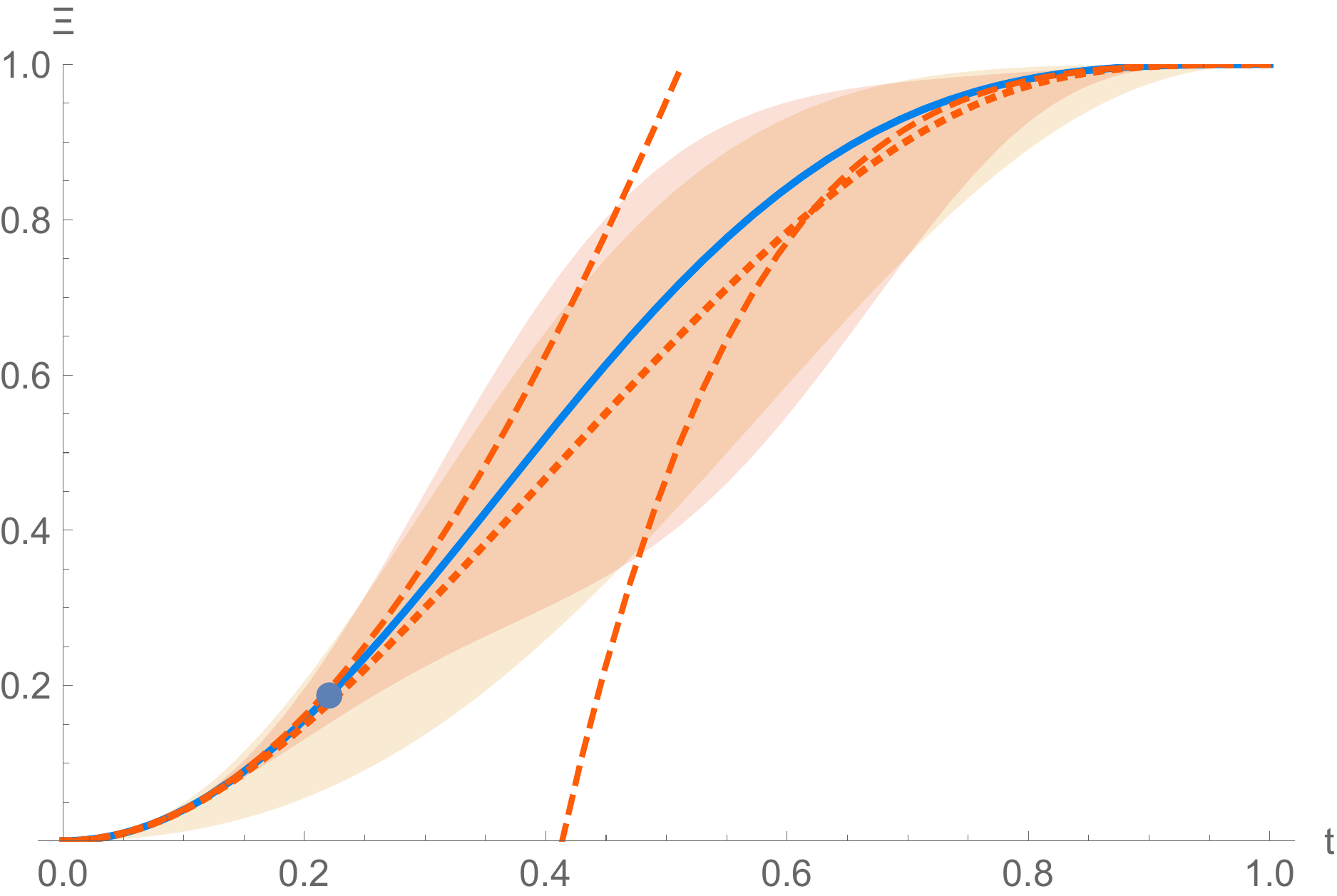}\\
\caption{
Results for the heterogeneous example of \cref{sec:example2} with $(\wtx_A,\wtx_B,\wtx_C) = (\nicefrac{1}{4},\nicefrac{3}{8},\nicefrac{1}{2})$ when no parametric form for $\wtx$ is specified, as in \cref{sec:polyParam} (left panels) or
when the events result from Poisson processes, as in \cref{sec:logParam} (right panels).
(Top) The parameterized probabilities (from bottom) $xa$, $xb$, and $xc$. 
(Bottom) The exact satisfiability $\Sat$ (solid curve) and interpolation (dotted curve) derived here along the parametric curve, along with truncated Taylor expansions for a depth-2 approximation at $\loc=0$ and depth-1 at $\loc=1$ (dashed lines). Notice that the Taylor series do not respect unitarity.
The lightly shaded region is bounded by the exact homogeneous solutions for $x_a$ and $x_c$;
the darker shading uses the monotonicity constraint with the approximate solution.
}
\label{fig:heteroCurve}
\end{center}
\end{figure}

Expressions for the elements of the matrix $\gamma$, defined in \cref{eq:gamma}, in terms of weights $a$, $b$, and $c$ defined in \cref{eq:hetero} for the corresponding probabilities $\wtx_A$,$\wtx_B$, and $\wtx_C$ are given in \cref{fig:gamma}.
The first row of $\vec{\gamma}$ comprises the elements of order $O(\wid^0)$.
It is thus independent of $a$, $b$, and $c$, and reduces to the Bernstein coefficients for the homogeneous case given in \cref{eq:exampleBeta}.
The terms given exactly by a depth 2 Inclusion-Exclusion expansion at $\loc=0$ and a depth-1 expansion at $\loc=1$ are indicated by the shading in \cref{fig:gamma} and in the table below. 
For ease of explication, here we consider only cases with $\wtx_B = (\wtx_A+\wtx_C)/2$, i.e. $a=-c=\pm1$ and $b=0$.
In this restricted example, the Bernstein transformations of the expressions in \cref{fig:gamma} reduce to:
\begin{equation}
\vec{\gamma}=
\left(
\begin{NiceArray}{>{\strut}cccccccc}%
[create-extra-nodes,left-margin,right-margin,
code-after = {\tikz \path [name suffix = -large, 
                                       fill = red!15,
                                       blend mode = multiply]
                          (1-1.north west)
                       |- (1-5.north east) 
                       |- (1-5.south east) 
                       |- (2-4.north east) 
                       |- (2-4.south east) 
                       |- (3-3.north east) 
                       |- (3-3.south east) 
                       |- (4-2.north east) 
                       |- (4-2.south east) 
                       |- (5-1.north east) 
                       |- (5-1.south east) 
                       |- (5-1.south west) 
                        |- (1-1.north west)  ;
                        \tikz \path [name suffix = -large, 
                                       fill = blue!15,
                                       blend mode = multiply]
                          (1-8.north east)
                       |- (1-6.north west) 
                       |- (1-6.south west) 
                       |- (4-6.south west) 
                       |- (5-5.north west) 
                       |- (5-5.south west) 
                       |- (6-4.north west) 
                       |- (6-4.south west) 
                       |- (7-3.north west) 
                       |- (7-3.south west) 
                       |- (8-2.north west) 
                       |- (8-2.south west) 
                       |- (8-8.south east) 
                       |- (1-8.north east) 
                       ; } ]
 0 & 0 & \frac{1}{21} & \frac{1}{5} & \frac{18}{35} & \frac{19}{21} & 1 & 1 \\[\matsep]
 0 & 0 & -\frac{2}{15} & -\frac{2}{5} & -\frac{17}{30} & -\frac{1}{6} & 0 & 0 \\[\matsep]
 0 & \frac{1}{8} & \frac{1}{4} & \frac{9}{40} & -\frac{1}{10} & 0 & 0 & 0 \\[\matsep]
 -\frac{1}{16} & -\frac{1}{16} & -\frac{1}{96} & \frac{3}{32} & 0 & 0 & 0 & 0 \\[\matsep]
 0 & -\frac{3}{256} & -\frac{3}{128} & 0 & 0 & 0 & 0 & 0 \\[\matsep]
 \frac{1}{512} & \frac{1}{512} & 0 & 0 & 0 & 0 & 0 & 0 \\[\matsep]
 0 & 0 & 0 & 0 & 0 & 0 & 0 & 0 \\[\matsep]
 0 & 0 & 0 & 0 & 0 & 0 & 0 & 0 
\end{NiceArray}
\right)
\label{eq:gammab0}
\end{equation}
The shading shows the coefficients that can be determined exactly using the truncated Taylor series.
Examining the third row, for example, illustrates that the rows of $\vec{\gamma}$ are not necessarily monotonic, and thus interpolations and bounds developed in \cref{sec:interp} are not immediately applicable.
Plots of $\Sat$ along the two parametric curves described in \cref{sec:polyParam} and \cref{sec:logParam} along with the bounds and interpolation available along the curve are shown in \cref{fig:heteroCurve} for the first case.
Of course, the heterogeneous case must also lie between the homogeneous cases for $x(t) =x_A(t)$ and $x_C(t)$, and these bounds are also  shown in the figure.

\Cref{tab:stats} displays results for several specific choices of $\vec{\wtx}$,
considering only the depth-$2$ Inclusion-Exclusion expansion for minimal struts and the depth-$1$ expansion for minimal cuts.
These specific cases were chosen to illustrate the effects of scaling and translating the probabilities or replacing them with their complements (i.e., transforming to the dual problem), and to compare the polynomial and logarithmic decompositions.

\section{Discussion}

We have described a novel approximation technique for mostly monotonic probabilistic satisfiability problems with event-specific probabilities.
Its first stage computes the weak- and strong-coupling perturbative expansions of statistical physics -- which are seen to be Inclusion-Exclusion expansions -- in terms of certain problem-specific minimal sets.
There may be up to $m=2^N$ of these sets; if they are not given explicitly, they may be sampled in at worst $O(N^2 \log N)$ steps for each of $S$ samples.
The expansions can be truncated at an arbitrary depth $\depth \le N$, producing  $\sum_{i=1}^D {M \choose i}$ terms at a cost of order $S^D$.
The truncated expansions separately have undesirable properties.
In particular, they do not respect the important constraints of monotonicity and unitarity.
Imposing unitarity allows them to be combined into a Bezier polynomial that provides a good approximation -- in absolute, if not necessarily relative, terms -- across the entire domain.
Imposing monotonicity allows the construction of problem-specific, tight upper and lower bounds on the approximation error due to truncation.
The bounds are tight in the sense that expressions with the same truncated expansions can be constructed whose satisfiabilities saturate both bounds, whereas any estimator that violates the bounds can be shown to be either non-monotonic or non-unitary.
It remains to be seen how sampling affects the quality of the bounds.

In this work, we have explored the weak- and strong-coupling perturbation series, which are Taylor series expansions around the distinguished points at $\loc = 0$ and $1$ and at $\wid=0$.
In the heterogeneous case, in addition to these points the point $(\loc,\wid)=(\nicefrac{1}{2},0)$ and the boundary $(\loc,\wid)=(\loc,\min(\loc,1-\loc))$ admit interesting simplifications.
The satisfiability at $(\nicefrac{1}{2},0)$ is simply related to the deterministic solution;
along the boundary, the probability of the most or least probable events $e_i$ is $1$ or $0$, respectively.
When $\wtx_i=0$, clauses that include $e_i$ can be ignored, because they never contribute to the satisfiability;
when $\wtx_i=1$, the event itself can be ignored, because it is always true.
In the dual representation, the same holds true for $\overline{\wtx}_i = 1 - \wtx_i$.
In principle, it is possible to develop perturbation series in $\wid$ around this point and the boundary segments.
This leads, in turn, to a step-wise renormalization approach that peels off the most and/or least probable events at each step.
Unfortunately, exploring these approaches is beyond the scope of this work.

The approximations and bounds developed here have been applied in the context of network reliability to a range of dynamical systems, including infectious disease transmission over a contact network \cite{NATH2018121}, crop pest movements over a commodity trade network \cite{Nath2019}, and the Ising model in the presence of an external field \cite{ren2016network}.
We hope that the synthesis presented here provides both a new perspective on some old problems and a tool that others will find useful in an even greater range of applications.

\begin{table}[tp]
\caption{(Top) The matrix of Taylor coefficients at $(\loc,\wid) = (0,0)$, $\vec{\alpha}$ defined in \cref{eq:heteroTaylor}; (bottom) the corresponding matrix $\vec{\overline{\alpha}}$ at (1,1) for the heterogenous example. Shading in the upper left (respectively, bottom right) indicates coefficients whose values are given exactly by the depth-2 perturbative expansion at $(0,0)$ (respectively, the depth-1 expansion at $(1,1)$).
Notice that the coefficients along the skew-diagonal match, and that the first row of each matrix, corresponding to $\wid=0$, agrees with the Taylor series for the homogeneous example, \cref{eq:exampleCoefs} and \cref{eq:exampleBarCoefs}.
}
\begin{center}
\resizebox{1.0\textwidth}{!}
{$\left(\,\begin{NiceArray}{>{\strut}cccccccc}%
[create-extra-nodes,left-margin,right-margin,
code-after = {\tikz \path [name suffix = -large, 
                                       fill = red!15,
                                       blend mode = multiply]
                          (1-1.north west)
                       |- (1-5.north east) 
                       |- (1-5.south east) 
                       |- (2-4.north east) 
                       |- (2-4.south east) 
                       |- (3-3.north east) 
                       |- (3-3.south east) 
                       |- (4-2.north east) 
                       |- (4-2.south east) 
                       |- (5-1.north east) 
                       |- (5-1.south east) 
                       |- (5-1.south west) 
                        |- (1-1.north west)  ;
                       } ]
 0 & 0 & 1 & 2 & 0 & -3 & 0 & 1 \\[\matsep]
 0 & 2 c & 2 (a+2 b) & 0 & -3 a-4 (2 b+c) & 0 & a+4 b+2 c & 0 \\[\matsep]
 c^2 & 2 b (2 a+b) & 0 & -2 (2 b+c) (2 a+2 b+c) & 0 & 2 a (2 b+c)+6 b^2+8 b c+c^2 & 0 & 0 \\[\matsep]
 2 a b^2 & 0 & -2 \left(a (2 b+c)^2+2 b \left(b^2+b c+c^2\right)\right) & 0 & a \left(6 b^2+8 b c+c^2\right)+4 b \left(b^2+3 b c+c^2\right) & 0 & 0 & 0 \\[\matsep]
 0 & -b \left(4 a \left(b^2+b c+c^2\right)+b^3+2 b c^2\right) & 0 & b \left(4 a \left(b^2+3 b c+c^2\right)+b \left(b^2+8 b c+6 c^2\right)\right) & 0 & 0 & 0 & 0 \\[\matsep]
 -a b^2 \left(b^2+2 c^2\right) & 0 & b^2 \left(a \left(b^2+8 b c+6 c^2\right)+2 b c (b+2 c)\right) & 0 & 0 & 0 & 0 & 0 \\[\matsep]
 0 & b^3 c (2 a (b+2 c)+b c) & 0 & 0 & 0 & 0 & 0 & 0 \\[\matsep]
 a b^4 c^2 & 0 & 0 & 0 & 0 & 0 & 0 & 0 \\[\matsep]
\end{NiceArray}\,\right)$}

\vspace{12pt}

\resizebox{1.0\textwidth}{!}
{$\left(\,\begin{NiceArray}{>{\strut}cccccccc}%
[create-extra-nodes,left-margin,right-margin,
code-after =  {\tikz \path [name suffix = -large, 
                                       fill = blue!15,
                                       blend mode = multiply]
                          (1-8.north east)
                       |- (1-6.north west) 
                       |- (1-6.south west) 
                       |- (4-6.south west) 
                       |- (5-5.north west) 
                       |- (5-5.south west) 
                       |- (6-4.north west) 
                       |- (6-4.south west) 
                       |- (7-3.north west) 
                       |- (7-3.south west) 
                       |- (8-2.north west) 
                       |- (8-2.south west) 
                       |- (8-8.south east) 
                       |- (1-8.north east) 
                       ; } ]
 1 & 0 & -2 & -7 & 20 & -18 & 7 & -1 \\[\matsep]
 0 & -2 (a+c) & a-16 b-6 c & 8 (a+6 b+3 c) & -2 (6 a+13 (2 b+c)) & 6 (a+4 b+2 c) & -a-2 (2 b+c) & 0 \\[\matsep]
 -2 a c & c (2 a+c)-8 b^2-16 b c & 4 \left(2 a (2 b+c)+9 b^2+14 b c+c^2\right) & -4 \left(4 a (2 b+c)+13 b^2+18 b c+2 c^2\right) & 5 \left(2 a (2 b+c)+6 b^2+8 b c+c^2\right) & -2 a (2 b+c)-6 b^2-8 b c-c^2 & 0 & 0 \\[\matsep]
 c \left(a c-8 b^2\right) & 8 b \left(a (b+2 c)+b^2+5 b c+c^2\right) & -4 \left(a \left(7 b^2+10 b c+c^2\right)+b \left(5 b^2+17 b c+5 c^2\right)\right) & 4 \left(a \left(6 b^2+8 b c+c^2\right)+4 b \left(b^2+3 b c+c^2\right)\right) & -a \left(6 b^2+8 b c+c^2\right)-4 b \left(b^2+3 b c+c^2\right) & 0 & 0 & 0 \\[\matsep]
 4 b^2 c (2 a+2 b+c) & -2 b \left(4 a \left(b^2+4 b c+c^2\right)+b \left(b^2+12 b c+8 c^2\right)\right) & 3 b \left(4 a \left(b^2+3 b c+c^2\right)+b \left(b^2+8 b c+6 c^2\right)\right) & -b \left(4 a \left(b^2+3 b c+c^2\right)+b \left(b^2+8 b c+6 c^2\right)\right) & 0 & 0 & 0 & 0 \\[\matsep]
 -2 b^2 c (2 a (2 b+c)+b (b+2 c)) & 2 b^2 \left(a \left(b^2+8 b c+6 c^2\right)+2 b c (b+2 c)\right) & -b^2 \left(a \left(b^2+8 b c+6 c^2\right)+2 b c (b+2 c)\right) & 0 & 0 & 0 & 0 & 0 \\[\matsep]
 b^3 c (2 a (b+2 c)+b c) & -b^3 c (2 a (b+2 c)+b c) & 0 & 0 & 0 & 0 & 0 & 0 \\[\matsep]
 -a b^4 c^2 & 0 & 0 & 0 & 0 & 0 & 0 & 0 \\[\matsep]
\end{NiceArray}\,\right)
$}
\label{fig:gamma}
\label{fig:gammaInterp}
\end{center}
\end{table}
\begin{table}
\begin{center}
\caption{Summary of approximation quality for several instances of $\vec{\wtx}$ in the heterogeneous example discussed in \cref{sec:example2}.
$\Sat$ is the true value of the satisfiability.
Estimates are constructed by interpolating between the left and right expansions for missing coefficients in $\vec{\theta}$.
The polynomial decomposition (labeled ``P'') uses the linear interpolation defined in \cref{sec:interp}; the logarithmic one (labeled ``L''), the logarithmic interpolation defined there.
The bounds are determined by monotonicity in $\vec{\theta}$.
Results for the polynomial parameterization are not shown in the last column, where $\wtx_N-\wtx_1>\nicefrac{1}{2}$, and it does not respect unitarity.
}
\begin{tabular}{rdddddd}
$(\wtx_A,\wtx_B,\wtx_C)$ 
 & \multicolumn{1}{c}{(0.22,0.31,0.39)}
 & \multicolumn{1}{c}{($\frac{1}{4}$,$\frac{3}{8}$,$\frac{1}{2}$)} 
 & \multicolumn{1}{c}{($\frac{1}{2}$,$\frac{3}{8}$,$\frac{1}{4}$)} 
 & \multicolumn{1}{c}{($\frac{1}{2}$,$\frac{5}{8}$,$\frac{3}{4}$)} 
 & \multicolumn{1}{c}{($\frac{1}{2}$,0.65,$\frac{3}{4}$)} 
 & \multicolumn{1}{c}{-}
 \\[\matsep]
$(\rate_A,\rate_B,\rate_C)$ 
 & \multicolumn{1}{c}{($\frac{1}{4}$,$\frac{3}{8}$,$\frac{1}{2}$)}
 & \multicolumn{1}{c}{-}
 & \multicolumn{1}{c}{-}
 & \multicolumn{1}{c}{-}
 & \multicolumn{1}{c}{($\ln 2$,$\frac{3}{2}\ln2$,$\ln$4)}
 & \multicolumn{1}{c}{($\frac{1}{4}$,$\frac{17}{8}$,4)}
 \\[\matsep]
 $\Sat({\cal E},\vec{\wtx})$ &  0.190 & 0.299 & 0.185 &  0.700 & 0.707 & 0.971\\[\matsep]
 \hline\\
estimate P& 0.185 & 0.288 & 0.190 & 0.666 & 0.668 & \multicolumn{1}{c}{-}\\
upper P& 0.258 & 0.413 & 0.343 & 0.878 & 0.879 & \multicolumn{1}{c}{-}\\
lower  P& 0.147 & 0.213 & 0.097 & 0.434 &0.438 &\multicolumn{1}{c}{-} \\
rel. diff. P& 0.022 & 0.037 & -0.025 & 0.049 &0.055 & \multicolumn{1}{c}{-}\\[\matsep]
 \hline\\
estimate L& 0.199  & \multicolumn{1}{c}{-} & \multicolumn{1}{c}{-} & \multicolumn{1}{c}{-} & 0.758 & 1.000\\
upper     L& 0.228  & \multicolumn{1}{c}{-} & \multicolumn{1}{c}{-} & \multicolumn{1}{c}{-}  & 0.867 & 1.000\\
lower      L& 0.162  & \multicolumn{1}{c}{-} & \multicolumn{1}{c}{-} & \multicolumn{1}{c}{-} & 0.366 & 0.362\\
rel. diff.   L& -0.051 & \multicolumn{1}{c}{-} & \multicolumn{1}{c}{-} & \multicolumn{1}{c}{-} &-0.072 & -0.029
\end{tabular}
\label{tab:stats}
\end{center}
\end{table}

\appendix

\section{Transforming a satisfiability problem into S-T reliability on a graph}
\label{app:SAT2Graph}
\begin{algorithmic}
\Require{${\cal L}$, a set of sets of integers in [1,\ldots,N]}, i.e., clauses in the expression's DNF form
\Function{SAT2Graph}{${\cal L}$}
\State $S \gets \{1,\ldots,N\}$; $T \gets \emptyset$;  ${V} \gets \{S, T\}$; $E \gets \emptyset$\Comment Initialize vertex and edge sets, $V$ and $E$
\State \Call{push}{$stack,(S,{\cal L})$}
\While {$stack$ is not empty} 
\State $(v,\ell) \gets$ \Call{pop}{$stack$}
\State $s \gets \cap_{c\in\ell}\, c$ \Comment $s$ contains only the events that appear in every clause in $\ell$
\For{$s_i \in s$} \Comment Add an incoming edge to $v$ and move to its source
\State $v' \gets v \setminus s_i$
\State $V \gets V \cup v'$; $E \gets E \cup (v', v)$; $Labels(v',v) \gets s_i$
\State  $v \gets v'$
\FFor{$\ell_j\in\ell$} $\ell_j\gets\ell_j\setminus s_i$\EndFFor \Comment Remove this event from further consideration
\EndFor
\State $s \gets \cup_{c\in\ell} \,c$ \Comment $s$ contains events that appear in any clause in $\ell$
\For{$s_i \in s$} \Comment Partition $\ell$ into clauses that do or don't contain $s_i$
\State $\ell' \gets \{c \in \ell | s_i \in c\}$; $\ell\gets\ell\setminus\ell'$
\If{$\ell' \neq \emptyset$} \Comment Add an incoming edge to $v$ and move to its source
\State $v' \gets \left( \cup_{c\in\ell'} \,c\right) \setminus s_i$
\State $V \gets V \cup v'$; $E \gets E \cup (v', v)$; $Labels(v',v) \gets s_i$
\FFor{$\ell_j \in \ell'$} $\ell_j \gets\ell_j \setminus s_i$ \EndFFor \Comment Remove this event from further consideration

\State \Call{push}{$stack,(v', \ell')$}
\EndIf
\EndFor
\EndWhile
\State \Return $(V,\,E,\,Labels)$
\EndFunction
\end{algorithmic}

\section{The expected value of $n_{k+1}$ given $n_k$}
\label{sec:expected}
Any given configuration at level $k+1$ is connected to $k+1$ configurations at level $k$, so the probability it is connected to any {\em specific} configuration is $\nicefrac{(k+1)}{{N\choose k}}$.
Hence, under the assumption that connections are independent, the probability that a configuration at level $k+1$ is not connected to any of $n_k$ configurations at level $k$ is $ \left[1 - \nicefrac{(k+1)}{{N \choose k}}\right]^{n_k}$, and 
the expected value of $n_{k+1}$ given $n_k$ is
\begin{equation}
\langle n_{k+1} \rangle = \left[1 - \left(1 - \frac{k+1}{{N \choose k}}\right)^{n_k}\right] {N \choose k+1} 
\label{eq:expStruts}
\end{equation}
In the limit $\beta_k \ll 1$, this reduces to
\begin{equation}
\langle n_{k+1} \rangle \xrightarrow[\beta_k \ll 1]{} (N-k) n_k,
\end{equation}
as it should.
In the example of \cref{sec:example}, the difference between this interpolation and linear interpolation is negligible.
In larger systems, with more undetermined coefficients, however, 
 the overlap is significant even for fairly small $\beta_k$. 
Further analysis is beyond the scope of this work.

\section*{Acknowledgments}
This material is based upon work supported by the following:
\begin{itemize}
\item the National Science Foundation under Grant No. CCF-1918656;
\item the National Institute of General Medical Sciences of the National Institutes of Health under Award Number R01GM109718;
\item the Defense Threat Reduction Agency (DTRA) under Contract No. HDTRA1-19-D-0007;
\item the United States Agency for International Development and the generous support of the American people through USAID Leader Award No. AID-OAA-L-15-00001;
\item and the Food and Agriculture Cyberinformatics and Tools Initiative Grant No. 2019-67021-29933 from the USDA National Institute of Food and Agriculture.
\end{itemize}
The content is solely the responsibility of the authors and does not necessarily represent the official views of any of these organizations.

\end{document}